\documentclass[aps,prd,amssymb,eqsecnum,showpacs,nofootinbib,floatfix,twocolumn,a4paper]{revtex4}%{revtex4-1}
\UseRawInputEncoding
\usepackage{amsmath,amssymb,yhmath,bm}
\usepackage{bbm}
\usepackage[all]{xy}
\usepackage{pstricks,pst-node,pst-text,pst-3d, psfrag,graphicx}
\usepackage{slashed}
\usepackage{color}
\usepackage{array}
\usepackage{tabulary}
\usepackage{rotating}
\newcommand{\be}{\begin{equation}}
\newcommand{\ee}{\end{equation}}
\newcommand{\bea}{\begin{eqnarray}}
\newcommand{\eea}{\end{eqnarray}}

\newcommand{\eqal}[1]{\begin{align}#1\end{align}}
\newcommand{\eqals}[1]{\begin{align*}#1\end{align*}}
\newcommand{\ft}[2]{{\textstyle\frac{#1}{#2}}}
 \newcommand{\ha}{{\hat a}}
\newcommand{\hb}{{\hat b}}
\newcommand{\N}{{\cal N}}
\newcommand{\Ni}{{({\cal N}^{-1})}}
\newcommand{\tN}{{\widetilde N}}

\begin{document}

\title{Hidden Kac-Moody Structures in the Fermionic Sector  \\ of Five-Dimensional  Supergravity}

\author{Thibault Damour\footnote{{\texttt Email : damour@ihes.fr}}$^1$ and Philippe  Spindel\footnote{{\texttt Email : philippe.spindel@umons.ac.be}}$^{2,3}$}

  \affiliation{
  $^1$Institut des Hautes \'Etudes Scientifiques, 91440 Bures-sur-Yvette , France.\\
$^2$Service de Physique de l'Univers, Champs et Gravitation, Universit\'e de Mons,\\ Facult\'e des Sciences,20, Place du Parc, B-7000 Mons, Belgium\\
$^3$ Service de Physique Th\'eorique,
Universit\'e Libre de Bruxelles\\
Bld du Triomphe CP225, 1050 Brussels, Belgium\\
}

\date{\today}

\begin{abstract}
We study the supersymmetric quantum dynamics of the cosmological models obtained 
by reducing $D=5$ supergravity to one timelike dimension. This consistent truncation has fourteen
bosonic degrees of freedom, while the quantization of the homogeneous gravitino field leads to a 
$2^{16}$--dimensional fermionic Hilbert space. We construct a consistent quantization of the model
in which the wave function of the Universe  is a $2^{16}$--component spinor %\textcolor{red}{of  Spin(24,8)}
depending on fourteen continuous coordinates, which satisfies eight Dirac-like wave equations
(supersymmetry constraints) and one Klein-Gordon-like equation (Hamiltonian constraint).
The fermionic part of the quantum Hamiltonian is built from operators 
that generate a $2^{16}$-dimensional representation 
of the (infinite-dimensional) maximally compact sub-algebra $K(G_2^{++})$
of the rank-4 hyperbolic Kac--Moody algebra $G_2^{++}$.  
The (quartic-in-fermions) squared-mass term $\widehat \mu^2$ entering the Klein-Gordon-like equation has several remarkable properties: (i) it commutes with the generators of $K(G_2^{++})$; and  (ii)
it is a quadratic polynomial in the fermion number $N_F \sim  \overline\Psi \Psi$, and a symplectic fermion bilinear
$C_F \sim \Psi C\Psi$. Some aspects of the structure of the solutions of our model are discussed, and notably
the Kac-Moody meaning of the operators describing the reflection of the wave function on the fermion-dependent
potential walls (``quantum fermionic Kac-Moody billiard").
\end{abstract}

\maketitle
%%%%%%%%%%%%%%%%%%%%%%%%%%%%%%%%%%%%%%%%%%%%%%%%

\section{Introduction}

 The discovery of a hidden $E_7$ symmetry of $N=8$ supergravity in $D=4$ \cite{Cremmer:1979up} has initiated 
 the search of hidden symmetries in supergravity, and superstring theories. The hidden symmetry algebra
 was more generally conjectured to be $E_{11-D}$  for maximal supergravity reduced to $D$ dimensions
 \cite{Julia80}, which implied  reaching the affine Kac-Moody algebra $E_9$ in $D=2$ \cite{Nicolai:1987kz},
 and, possibly, the hyperbolic Kac-Moody algebra $E_{10}$ when reducing to one timelike direction \cite{Mizoguchi:1997si}. [See Ref. \cite{Kac} for the definition and basic structure of infinite-dimensional Kac-Moody algebras.]
 The possible existence of a hidden, mother $E_{11}$ structure has been suggested in \cite{West:2001as}.
% see Ref. \cite{Bossard:2021ebg} for recent work along a   related ``exceptional" line.
 
 A new angle on the possible relevance of $E_{10}$ came from studies of the chaotic behavior,
 \`a la Belinskii-Khalatnikov-Lifshitz \cite{Belinsky:1970ew,Henneaux:2007ej,Belinski:2017fas}, of 
 generic solutions of maximal supergravity near a (spacelike) cosmological singularity 
 \cite{Damour:2000hv,Damour:2002cu,Damour:2002et}. These studies highlighted the role of the gravitino in the
 implementation of hidden hyperbolic Kac-Moody structures \cite{Damour:2005zs,de Buyl:2005mt,Damour:2006xu}.
 The gravitino enters the game as a representation of the algebra $K(E_{10})$, defined
 as the (formal) maximally compact subalgebra of $E_{10}$, namely the subalgebra fixed under the
 Chevalley involution. [We use the maximally split real forms of the considered hyperbolic Kac-Moody algebras,
 and the corresponding real Chevalley involution.]
 The existence of {\it finite-dimensional} spinorial representations of (infinite-dimensional) involutory subalgebras
 of hyperbolic Kac-Moody algebras discovered through such supergravity-based works 
 \cite{Damour:2005zs,de Buyl:2005mt,Damour:2006xu,Damour:2009zc} was extended in several directions 
 \cite{Kleinschmidt:2013eka,Damour:2017cpi,Kleinschmidt:2018hdr}, and notably from a
 mathematical point of view \cite{Koehl,Koehl2,Koehl3,Kleinschmidt:2021agj}.
 
 Most studies, however, only considered the gravitino dynamics at lowest order, where the gravitino
 can be treated as a classical, Grassmanian field, undergoing a fermionic analog \cite{Damour:2009zc}
 of the bosonic billiard dynamics.  The compatibility of Kac-Moody structures
 with the fully nonlinear gravitino dynamics (involving up to quartic-in-fermions terms in the Hamiltonian)
 has only been explored so far within the simpler setting of the reduction of $N=1$, $D=4$ supergravity 
 to homogeneous cosmological models of the Bianchi IX type 
 \cite{Damour:2013eua,Damour:2014cba,Damour:2017cpi}. In this setting, the relevant
 hyperbolic Kac-Moody structures are not $E_{10}$ and $K(E_{10})$, but a rank-3 hyperbolic subalgebra 
 of $E_{10}$ called $AE_3$, and its maximally compact subalgebra $K(AE_3)$. In these studies the
 gravitino is treated as a fully nonlinear  quantum field (depending only on time). The results of
 Refs. \cite{Damour:2013eua,Damour:2014cba,Damour:2017cpi}  have deepened
 the significance of hidden Kac-Moody structures by showing, in particular, that:
 (i) the quartic-in-fermion contribution to the quantum Hamiltonian is invariant under the three generators 
 $\widehat J_{\alpha_1}$, $\widehat J_{\alpha_2}$, $\widehat J_{\alpha_3}$ of $K(AE_3)$ 
 (which are associated with the three
 simple roots $\alpha_1,\alpha_2,\alpha_3$  of $AE_3$);
 and
 (ii) the quantum dynamics of the gravitino near the singularity can be described as a sequence 
 of free motions interrupted by reflections on three Toda-like potential walls corresponding to the three
 simple roots of $AE_3$.  Each such reflection is described (in the short-wavelength limit) by
 the corresponding quantum reflection operator (with $i=1,2,3$)
 \be\label{Ralphaq}
{\widehat {\mathcal R}}_{\alpha_i} = e^{ i  \frac{\pi}{2}  \widehat J_{\alpha_i}} \,.
\ee
 In addition, the latter reflection operators satisfy a generalized version of the Coxeter relations
 satisfied by usual hyperplane reflection operators.

The aim of the present work is to extend the work of Refs.  \cite{Damour:2013eua,Damour:2014cba,Damour:2017cpi}
to the case of pure supergravity in $D=5$, 
as a step towards understanding the nonlinear aspects of
fermions in $D=11$ supergravity. We recall that pure $D=5$ supergravity  (with eight supercharges)
exhibits some similarity with $D=11$ supergravity \cite{Mizoguchi:1998wv}. It is therefore interesting to study
the compatibility of Kac-Moody structures with the fully nonlinear gravitino dynamics within the simpler setting
of $D=5$  supergravity. Previous works have indicated that, in this case, the relevant hyperbolic Kac-Moody
algebra behind the bosonic dynamics was the rank-4 hyperbolic extension of $G_2$, which we will denote
as $G_2^{++}$  \cite{Damour:2002fz,Mizoguchi:2005zf}. [ Contrary to  $E_{10}$  (but similarly to $AE_3$) 
the hyperbolic Kac-Moody  $G_2^{++}$ is non-simply laced.]
We therefore expect that the gravitino
will enter as a representation of the subalgebra $K(G_2^{++}) \subset G_2^{++}$, fixed under
the Chevalley involution. We will indeed find that the $D=5$ supergravity fermion couplings define
a consistent finite-dimensional vector-spinor representation of $K(G_2^{++})$  (of the type
defined in Ref. \cite{Kleinschmidt:2018hdr}), and we shall
prove that analogs of the results found for the $K(AE_3)$ structure of $D=4$ supergravity cosmological
models hold for the simplest homogeneous cosmological models of $D=5$ supergravity (where 
all fields are taken to depend only on time). In particular, 
the quartic-in-fermion contribution to the quantum Hamiltonian will be shown to be invariant under the four generators 
 $\widehat J_{\alpha_1}$, $\widehat J_{\alpha_2}$, $\widehat J_{\alpha_3}$, $\widehat J_{\alpha_4}$,
  of $K(G_2^{++})$,   associated with the four
 simple roots $\alpha_1,\alpha_2,\alpha_3,\alpha_4$  of $G_2^{++}$.

%%%%%
\section{Classical Lagrangian formulation}

We take as starting point the second-order action of the pure supergravity theory in $D=5$, as given in (the corrected
version of) Ref. \cite{CN}. In this formulation the gravitino is described by a (complex) Dirac vector-spinor $\psi_\mu$.
[This is equivalent to the alternative formulation using a doublet of symplectic Majorana vector-spinors \cite{Cremmer:1980gs}.]
We follow the normalization and notation of Ref. \cite{CN}, notably for the Levi-Civita connection
$\ring\omega_{\mu\,{\hat   \alpha} {\hat   \beta}}= - \ring\omega_{\mu\, {\hat   \beta} {\hat   \alpha}}$. Here, $\mu$ is
a five-dimensional coordinate index, while hatted indices are frame indices with respect to a local Lorentz frame $ e_{{\hat   \alpha}}^{\mu}$, 
with associated coframe $\theta^{\hat   \alpha}_\mu$ 
($e_{{\hat   \alpha}}^{\mu} \, \theta^{\hat   \beta}_\mu =\delta_{{\hat   \alpha}}^{\hat   \beta}$).  The Levi-Civita connection (with one coordinate index, $\mu$, and two frame indices) is defined as
\be
\ring\omega_{\mu\,{\hat   \alpha} {\hat   \beta}}\equiv \eta_{{\hat   \alpha}{\hat   \gamma} } \,\ring\omega_{\mu \ \, \hat   \beta}^{\ \hat   \gamma}= - \ring\omega_{\mu\, {\hat   \beta}{\hat   \alpha} }\, ,
\ee
with
\be
\ring\omega_{\mu \ \, \hat   \beta}^{\ \hat   \alpha}\equiv + \theta^{\hat   \alpha}_\nu \left(\partial_\mu\, e_{{\hat   \beta}}^{\nu}+\Gamma_{\mu\sigma}^\nu\,e_{\hat   \beta}^\sigma\right)\,, %= -\ring\omega_{\mu  \hat   \beta}^{\ \ \,\hat   \alpha}\,. 
\ee 
where $\Gamma_{\mu\sigma}^\nu$ denote the usual Christoffel symbols of $g_{\mu\nu}$.

The covariant derivatives of the frame components of a vector, and of a vector-spinor, are respectively given by (when using frame indices)
\eqal{
\nabla_\mu \,V^{\hat   \alpha}=&\partial_\mu V^{\hat   \alpha}+\ring\omega_{\mu\phantom{{\hat   \alpha}}{\hat   \beta}}^{\phantom{\alpha}{\hat   \alpha}}\,V^{\hat   \beta}\,, \\
{\cal D}_\lambda[\ring \omega] \psi_{\hat   \mu}=&\partial_\lambda\psi_{\hat   \mu}+\ring\omega_{\lambda{\hat   \mu}}^{\phantom{\alpha{\hat   \mu}}{\hat   \nu}}\psi_{\hat   \nu}+\ft 14\, \ring\omega_\lambda^{\phantom{\alpha}{{\hat   \rho}}{{{\hat   \sigma}}}}\gamma_{{{\hat   \rho}}{{{\hat   \sigma}}}}\psi_{\hat   \mu}\,.}
As we use here a mostly positive signature, we had to adapt the results of Ref. \cite{CN} (which used a mostly negative signature). 
For instance, we replaced their gamma matrices as follows: $ \Gamma_{CN}^{\hat \mu} \mapsto -i\,\gamma^{\hat \mu}$,
$ \Gamma^{CN}_{\hat \mu} \mapsto +i\,\gamma_{\hat\mu}$. Our gamma matrices satisfy 
$\gamma_{\hat \mu} \gamma_{\hat \nu} + \gamma_{\hat \nu} \gamma_{\hat \mu}= 2 \eta_{\hat \mu \hat \nu}$
with $ \eta_{\hat \mu \hat \nu}= {\rm diag}(-1,+1,+1,+1,+1)$.

Our sign convention
for the covariant components of the antisymmetric Levi-Civita tensor is $\eta_{\alpha \beta \gamma \delta \epsilon}= 
 \sqrt{|g|}  \varepsilon_{\alpha \beta \gamma \delta \epsilon}$ with $\varepsilon_{01234}=+1$. The antisymmetrized product of five 
 gamma matrices is proportional to the identity matrix and we use (following \cite{CN}) a representation where $\gamma_{\mu\nu\rho\sigma\tau}=-i\,\eta_{\mu\nu\rho\sigma\tau}$, {\it i.e.}, $\gamma_{01234}=-i\, \sqrt{\vert g\vert}$,  or $\gamma_{\hat 0\hat 1\hat 2\hat 3\hat 4}=-i$,
so that $\gamma^{\hat 0\hat 1\hat 2\hat 3\hat 4}=+i$.
We define the Dirac conjugate as
\be
{\overline\Psi}\equiv \,\Psi^\dagger \beta\,,
\ee
with the $\beta$ matrix defined such that $ \beta \gamma_{\hat \mu} \beta^{-1}= - \gamma_{\hat \mu}^{\dagger}$.
We take a representation of the (positive-signature) gamma matrices 
where $ \gamma^{\hat  0}$ is anti-hermitian, while  the $\gamma^{\hat  i}$'s are hermitian, and choose
\be
\beta \equiv + i\,\gamma_{\hat  0}= - i\, \gamma^{\hat  0}\,.
\ee
Note that $\beta$ is hermitian
and unipotent:
\be
\beta^{\dagger}= \beta \; ; \; \beta^2=1\,.
\ee
%\be
%{\overline\Psi}\equiv + i\,\Psi^\dagger \gamma_{\hat  0}= - i\,\Psi^\dagger \gamma^{\hat  0}\,.
%\ee
The action reads $S = \int d^5x \mathcal L$, with Lagrangian density $\mathcal L=e\,L$
 (with $e = \det e^{{\hat   \alpha}}_{\mu}= \sqrt{|g|}$), and a
second-order Lagrangian $L$ given (in units where $4 \pi G_5=1$) by
\eqal{
 L=&\ft1{4} R(\ring\omega) -\ft1 4 
{ F}_{\mu \nu}{ F}^{\mu \nu} +\,\ft{1}{6\sqrt{3}}   \eta ^{\mu \nu \lambda \rho \sigma }  
A_\mu  F_{\nu\lambda }  F_{\rho \sigma }\nonumber{}\\
&+\,\ft1{ 2}\Big( {{\overline\psi}}_\mu \gamma^{\mu \nu \rho}\mathcal{D}_\nu( {\ring\omega})  
\psi_\rho  -\overline{\mathcal{D}_\nu(\ring\omega)\psi_\mu} \gamma^{\mu \nu \rho} \psi_\rho\Big)\nonumber{} \\
  &-i\ft{\sqrt{3}}{4} \Big({{\overline\psi}}_\mu \gamma^{\mu\nu\rho\sigma}\psi_\nu+ {\overline\psi}^{\rho}\psi^{ \sigma}-{\overline\psi}^{ \sigma}\psi^\rho \Big)  F_{\rho\sigma} \nonumber \\
  &  +\bar{\psi}_{[\mu}\gamma^\mu\psi_{\alpha]}\bar{\psi}^{[\nu}\gamma_\nu\psi^{\alpha]}\nonumber 
-\,\ft{1}{2}\bar{\psi}_{[\mu}\gamma_{\vert \nu\vert}\psi_{\rho]}\bar{\psi}^{[\mu}\gamma^{\vert \rho\vert}\psi^{\nu]} \nonumber{} \\
&-\ft{1}{4}\bar{\psi}_{[\mu}\gamma^\nu\psi_{\rho]}\bar{\psi}^{[\mu}\gamma_\nu\psi^{\rho]} \nonumber{} 
+\ft{1}{4 }\bar{\psi}_\mu\psi_\nu\bar{\psi}_\rho\gamma^{\mu\nu\rho\sigma}\psi_\sigma \\&
+\ft{3}{8 }(\bar{\psi}_\mu\psi_\nu-\bar{\psi}_\nu\psi_\mu)\bar{\psi}^\mu\psi^\nu \,.\label{Lagtot}}
A consistent truncation of this theory consists in considering a  (Bianchi-I) five-dimensional
``minisuperspace'' cosmological model where all the fields ($g_{\mu \nu}$, $A_\mu$, $\psi_\mu$) 
depend only on time, without any spatial dependence. More precisely, we consider a model where 
the four-dimensional space is toroidally compactified (with  $ 0\leq x^i\leq 1$, $i=1,2,3,4$),
so that supergravity reduces to a kind of supersymmetric quantum mechanical model for the 
zero modes $g_{\mu \nu}(t)$, $A_\mu(t)$, and $\psi_\mu(t)$. 
The metric is written as
\be
ds^2=-N(t)^2dt^2+h_{ij}(t)(dx^i+N^i(t)dt)(dx^j+N^j(t)dt)\,.
\ee
The time component $A_0$ of the $A_\mu$ field drops out of the dynamics (the associated Gauss
constraint being identically zero). Similarly, the shift vector $N^i(t)$ drops out of the dynamics
(its associated momentum constraint vanishing identically). We henceforth set both $A_0$
and $N^i$ to zero. The only constraints that will remain in our cosmological dynamics are:
(i) the Hamiltonian constraint (associated with the lapse function $N(t)$); and (ii) the supersymmetry
constraint (associated with $\psi_0(t)$). 

As in our previous work dealing with a supersymmetric Bianchi-IX model in $D=4$ \cite{Damour:2014cba}, we shall avoid
the presence of constraints linked to local Lorentz rotations by using a local frame that is algebraically
defined in terms of the metric components $g_{\mu \nu}$. We use ($i,j,k=1,2,3,4$)
\eqal{&\theta^{{\hat  0} }=N\,dt\qquad,\qquad \theta^{{\hat  a}}=\theta^{{\hat a}}_i dx^i \,,\nonumber{}\\
&e_{\hat  0} =\ft 1N\partial_t \qquad,\qquad e_{\hat  a}= e^i_{\hat  a}\partial_i \,,}
where $\theta_j^{\hat  a}\,e_{\hat  a}^i=\delta^i_j$.
Previous work on the approach to cosmological 
singularities \cite{Damour:2002et} has emphasized the usefulness of parametrizing the gravitational degrees of freedom
by means of an Iwasawa decomposition of the spatial co-frame $\theta^{{\hat a}}_i $. This means encoding
the ten independent components of the spatial metric $h_{ij}$ by means of four diagonal logarithmic scale
factors $\exp(-{\beta^{\hat a}})$ and six off-diagonal variable ${n^{\hat a}}_i$ (with ${\hat a} <i$), 
defined so that
\be
\theta^{\hat a}_i=e^{-\beta^{\hat a}}({\cal N})^{\hat a}_{\ i}\quad,\quad e_{\hat  a}^i=e^{\beta^{\hat a}}({\cal N}^{-1})^i_{\ {\hat a}}\,,
\ee
where $\cal N$ is an upper triangular, unipotent matrix, namely
\be
{\cal N}= \big({\cal N}^{\hat a}_{\ \  i}\big)= \big(\delta^{\hat a}_{\ \  i} + n^{\hat a}_{\ \  i}\big)=
\left(
\begin{array}{cccc}
1  & { n^1}_{ 2}  &{ n^1}_{3}  &{ n^1}_{4} \nonumber{}\\
 0 &1   &{n^2}_{3}  & {n^2}_{4}\nonumber{}\\
 0 &0   & 1  &{n^3}_{4}\nonumber{}\\
 0&0&0&1
\end{array}
\right).
\ee
Note that the inverse matrix $\Ni^i_{\  \ha}$ is also an unipotent upper-triangular matrix.

As a consequence the spatial metric $h_{ij}$ reads
\bea \label{iwasawa}
h_{ij} &=& \sum_\ha e^{-2\beta^{ \hat a}} \N^\ha_{\ \  i} \,\N^\ha_{\ \ j} \,, \,{\it i.e.},\nonumber\\
  (h_{ij})&=&  {\cal N}^T{\cal A}^2{\cal N} \, ; \, {\rm with} \; {\cal A}\equiv{\rm diag}\{e^{-\beta^{ \hat a}} \}\,.
\eea
It is convenient to use as basic variables in the Lagrangian formulation the quantities
\be
\beta^{ \hat a} \,;\, {n^{\hat a}}_{ i}\, ({\rm with} \;\hat a <i);\, B_{\hat a}\,;\, \Psi_{\hat a}\,; \, {\widetilde N}\,,
\ee
where we defined (to replace $A_i$, $\psi_i$ and $N$)
\be
 B_\ha \equiv A_i({\cal N}^{-1})^i_{\ \ha}\,;\, \Psi_\ha\equiv e^{-\frac12 \sigma_\beta} \psi_\ha \,;\,{\widetilde N}\equiv N\,e^{+ \sigma_\beta}\,,
\ee
with $\sigma_\beta\equiv \sum_{\ha=1}^4 \beta^\ha$.

The Lagrangian density  $\mathcal L=e\,L$ then decomposes into
\be \label{Ldecomp}
\mathcal L= \mathcal L_{R} + \mathcal L_{F^2}+ \mathcal L_{RS}+ \mathcal L_{F \Psi^2}+ \mathcal L_{\Psi^4}\,,
\ee
where $ \mathcal L_{R} =\frac{e}{4} R(\ring\omega)$ corresponds to the first (Einstein-Hilbert) term in Eq. \eqref{Lagtot},
$\mathcal L_{F^2} = -\frac{e}4 { F}_{\mu \nu}{ F}^{\mu \nu} $ to the second (Maxwell) term, 
$\mathcal L_{RS}$ to the Rarita-Schwinger term on the second line, $ \mathcal L_{F \Psi^2}$ to
the $ \bar \psi \psi F$ coupling on the third line, and where $\mathcal L_{\Psi^4}$ corresponds
to all the remaining terms, which are quartic in $\psi$. [The Chern-Simons term $ A\wedge  F \wedge F$
on the first line vanishes, as well as its variation.] 
In our units (where $4 \pi G_5=1$ and $\int d^4 x=1$), we can
 consider $\mathcal L$ as the total Lagrangian of a supersymmetric
quantum mechanical model, with corresponding action $S= \int dt \mathcal L$.

The explicit expressions of the various terms in $\mathcal L$, Eq. \eqref{Ldecomp}, are as follows.
The Einstein term reads (henceforth, we cease to systematically put hats on the frame indices $a=\ha$)
\be
\mathcal L_{R} = \frac1{4 \widetilde N} \left( G_{ab}\dot\beta^{ a} \dot\beta^{ b}+\ft12 \sum_{a<b}e^{+2({\beta}^{  b}-{\beta}^{ a})} \left( W_{{\hat a} {\hat b}}\right)^2 \right)\,,
\ee
where the quadratic form $G_{ab}$
defining the kinetic terms of the logarithmic scale factors $\beta^a$ is defined as
\be \label{Gab}
G_{ab} \dot\beta^{ a} \dot\beta^{ b} \equiv \sum_a(\dot\beta^{ a})^2-(\sum_a\dot\beta^{ a})^2\,,
\ee

and where we defined (for ${\hat a}< {\hat b}$)
\be \label{Wab}
W_{{\hat a} {\hat b}} \equiv {W^{\hat a}}_{\ {\hat b}}\equiv\sum_{\ha<i \leq\hb}\dot n^{ \ha}_{\ i}({\cal N}^{-1})^i_{\ \hat b}\,.
\ee

The Maxwell kinetic term reads
\be
\mathcal L_{F^2} = \frac1{2 \widetilde N} \sum_a e^{2\beta^{ a}}\, E_a^2\,,
\ee
where $E_a$ denotes the electric-field variable
\bea
{ E}_a &\equiv& \sum_i({\cal N}^{-1})^i_{\hat  a} \,F_{ti}\nonumber{}\\
&=&\dot B_{ \ha} +\sum_{i=2}^\ha\sum_{\hb=1}^{i-1}B_{ \hb}\, \dot{n}^{\hb}_i({\cal N}^{-1})^i_\ha
\eea
The Rarita-Schwinger term reads
\be
 \mathcal L_{RS} =  \ft i2 \,G_{ab}\left({  \Phi}^{\dagger a} \dot\Phi^b-\dot{\Phi}^{ \dagger a}\Phi^b\right)
+\ft N2\, {\widetilde Q}^{{\hat   \nu}{\hat   \alpha}{\hat   \beta}}\ring\omega_{{\hat   \nu}{\hat   \alpha}{\hat   \beta} }\,.
\ee
Here we replaced the rescaled gravitino $\Psi^a$ by the useful vector-spinor variable \cite{Damour:2009zc}
\be \label{defPhi}
\Phi^a \equiv \gamma^\ha \Psi^\ha \; ; \; ({\rm no \; sum \; on}\; \ha)\,,
\ee
while the second term involves the contraction between the Fermion bilinear
\begin{widetext}
\eqal{{\widetilde Q}^{{\hat   \nu}{\hat   \alpha}{\hat   \beta}} &=
\ft12\Big(\overline \Psi_{\hat \mu}\gamma^{\hat \mu\hat \nu\hat \alpha}\Psi^\beta-\overline \Psi_{\hat \mu}\gamma^{\hat \mu\hat \nu\hat \beta}\Psi^\alpha-\overline \Psi^{\hat \beta}\gamma^{\hat \alpha\hat \nu\hat \mu }\Psi_{\hat \mu}+\overline \Psi^{\hat \alpha}\gamma^{\hat \beta\hat \nu\hat \mu }\Psi_{\hat \mu}\Big) 
+\ft12\overline\Psi_{\hat \mu}\gamma^{\hat \mu\hat \nu\hat \rho\hat \alpha\hat \beta}\Psi_{\hat \rho}\nonumber\\
&+\ft12\Big(\overline\Psi_{\hat \mu}\big(\gamma^{\hat \mu}(\eta^{\hat \rho\hat \alpha}\eta^{\hat \nu\hat\beta}-\eta^{\hat \nu\hat \alpha}\eta^{\hat \rho\hat\beta})+\gamma^{\hat \nu}(\eta^{\hat \mu\hat \alpha}\eta^{\hat \rho\hat\beta}-\eta^{\hat \rho\hat \alpha}\eta^{\hat \mu\hat\beta})+\gamma^{\hat \rho}(\eta^{\hat \nu\hat \alpha}\eta^{\hat \mu\hat\beta}-\eta^{\hat \mu\hat \alpha}\eta^{\hat \nu\hat\beta})\big)\Psi_{\hat \rho}\Big) \,, } 
\end{widetext}
and the Levi-Civita spin-connection, whose only nonvanishing (frame) components are
\eqal{
\ring{\omega}_{{\hat  0}  {\hat a}{\hat b}}=&-\ft 1{2\,N}\left(e^{-({\beta}^{a}-{\beta}^{ b})}\,W_{{\hat a}{\hat b}}-e^{-({\beta}^{  b}-{\beta}^{ a})}\,W_{{\hat b}{\hat a}}\right) \,, \nonumber{}\\
\ring\omega_{{\hat b}{\hat  0}   {\hat a}}=& \ft 1N \left( \dot\beta^{ a}\,\delta_{{\hat a}{\hat b}}-\frac12(e^{-({\beta}^{ a}-{\beta}^{ b})}\,W_{{\hat a}{\hat b}}+e^{-({\beta}^{  b}-{\beta}^{ a})}\,W_{{\hat b}{\hat a}})\right)\nonumber{}\\
=&\ring\omega_{{\hat a}{\hat  0}   {\hat b}} \, .}
Here the quantities $W_{{\hat a} {\hat b}}$ (which are essentially the time derivatives of $n^\ha_{\ i}$)
were defined in Eq. \eqref{Wab} above.
Note that $W_{{\hat a} {\hat b}}$ vanishes if $\hb \leq \ha$, so that
the non vanishing contributions to $\ring{\omega}$ are all multiplied by a factor of the type 
$e^{+({\beta}^{  b}-{\beta}^{ a})}$ with $ b > a$.

The $ \bar \psi \psi F$ coupling term, $ \mathcal L_{F \Psi^2}$, reads
\be
 \mathcal L_{F \Psi^2} = - i \frac{\sqrt{3}}{2} \sum_\ha e^{\beta^\ha} X^{\hat 0 \ha} E_\ha \,,
\ee
where
\be
X^{\hat 0 \ha} \equiv i \, \eta^{b c d \hat 0 a} \bar \Psi_b \gamma_c \Psi_d +  \bar \Psi^{\hat 0} \Psi^a-  \bar \Psi^a \Psi^{\hat 0}\,.
\ee
At this stage, we see that  the Lagrangian is the sum of four types of terms: (i) the kinetic terms for
the bosonic variables $\beta^{ \hat a} \,;\, {n^{\hat a}}_{ i}\,;\, B_{\hat a}$, namely,
\bea
\mathcal L_{\rm kin \, b}&=&  \frac1{ \widetilde N} \left( \frac14 G_{ab}\dot\beta^{ a} \dot\beta^{ b}   + \frac18 \sum_{a<b}e^{+2({\beta}^{  b}-{\beta}^{ a})} \left( W_{{\hat a} {\hat b}}\right)^2 \right. \nonumber\\ 
 &+&\left.   \frac12 \sum_a e^{2\beta^{ a}}\, E_a^2 \right);
\eea
(ii) the kinetic terms for the fermionic variables $ \Psi_{\hat a}$, namely
\be
\mathcal L_{\rm kin \, f}= \ft i2 \,G_{ab}\left({  \Phi}^{\dagger a} \dot\Phi^b-\dot{\Phi}^{ \dagger a}\Phi^b\right);
\ee
(iii) the couplings between the bosonic velocity variables $\dot \beta^a, W_{ab}, E_a$ and 
corresponding fermion bilinears, $Q^a_{\dot \beta}(\bar \Psi, \Psi), Q^{ab}_{W}(\bar \Psi, \Psi), 
Q^a_E(\bar \Psi, \Psi)$, of the form
\bea
\mathcal L_{\dot{q} \Psi^2}&=& \sum_a \dot \beta^a Q^a_{\dot \beta}(\bar \Psi, \Psi) + \sum_{a<b} e^{{\beta}^{  b}-{\beta}^{ a}} W_{ab}Q^{ab}_{W}(\bar \Psi, \Psi) \nonumber\\
&+& \sum_a  e^{\beta^{ a}}\, E_a Q^a_E(\bar \Psi, \Psi)\,,
\eea
with, for instance
\bea
 Q^a_{\dot \beta}(\bar \Psi, \Psi) &=& {\widetilde Q}^{{\hat a}{\hat  0}{\hat a}} \nonumber\\
&=&\ft i2\sum_{\hat m\neq \hat a} \big(\Psi ^{\dagger \hat m }\gamma^{\hat m\hat a}\Psi^{\hat a}-\Psi^{\dagger \hat a}\gamma^{\hat a\hat m}\Psi^{\hat  m} \nonumber\\
&+& \Psi ^{\dagger \hat  0}\gamma^{\hat 0\hat m}\Psi^{\hat m}+\Psi^{\dagger \hat m}\gamma^{\hat m\hat 0}\Psi^{\hat  0}\big) \nonumber \\
&=&\ft i2\sum_{\hat m\neq \hat a}\big (\Phi ^{\dagger \hat m }\Phi^{\hat a}-\Phi^{\dagger \hat a} \Phi^{\hat  m} \nonumber\\
&+& \Psi ^{\dagger \hat  0}\gamma^{\hat 0}\Phi^{\hat m}+\Phi^{\dagger \hat m}\gamma^{ \hat 0}\Psi^{\hat  0}\big ) ;
\eea
and, finally, (iv) the terms quartic in the fermions that entered the original Lagrangian, Eq.\eqref{Lagtot}, namely
\be
\mathcal L_{\Psi^4} =  {\widetilde N} L_{\Psi^4} \,,
\ee
with
\bea \label{LPsi4}
 L_{\Psi^4} &=&  \bar{\Psi}_{[\mu}\gamma^\mu\Psi_{\alpha]}\bar{\Psi}^{[\nu}\gamma_\nu\Psi^{\alpha]}\nonumber 
-\,\ft{1}{2}\bar{\Psi}_{[\mu}\gamma_{\vert \nu\vert}\Psi_{\rho]}\bar{\Psi}^{[\mu}\gamma^{\vert \rho\vert}\Psi^{\nu]} \nonumber{} \\
&&-\ft{1}{4}\bar{\Psi}_{[\mu}\gamma^\nu\Psi_{\rho]}\bar{\Psi}^{[\mu}\gamma_\nu\Psi^{\rho]} \nonumber{} 
+\ft{1}{4 }\bar{\Psi}_\mu\Psi_\nu\bar{\Psi}_\rho\gamma^{\mu\nu\rho\sigma}\Psi_\sigma \\&&
+\ft{3}{8 }(\bar{\Psi}_\mu\Psi_\nu-\bar{\Psi}_\nu\Psi_\mu)\bar{\Psi}^\mu\Psi^\nu \,.
\eea

%%%%%%%%
\section{Classical Hamiltonian formulation}

We have seen in the previous section that the Lagrangian had a structure of the type
\be
\mathcal L=  \ft i2 \,G_{ab}\left({  \Phi}^{\dagger a} \dot\Phi^b-\dot{\Phi}^{ \dagger a}\Phi^b\right)+\frac{1}{2 \tN} \dot q^k g_{kl}\dot q^l+\mathcal Q_k(\Psi)\dot q^k + \tN  L_{\Psi^4}\,,
\ee
where $q^k$ denote the bosonic variables, $\beta^{ \hat a} \,;\, {n^{\hat a}}_{ i}\,;\, B_{\hat a}$,
 where the  $\mathcal Q_k$'s are bilinear in the fermions (and depend on the bosonic variables,
notably through various exponential factors $e^{{\beta}^{  b}-{\beta}^{ a}},  e^{{\beta}^{ a}} $),
and where the term quartic in the fermions, $ L_{\Psi^4}$, is given by Eq.  \eqref{LPsi4}. We recall
that  $\Phi^a$ denote the redefined version \eqref{defPhi} of the gravitino variables  $\Psi^a$.

Passing to the corresponding Hamiltonian formulation, in terms of the bosonic momenta,
\be
p_k = \frac{\partial \mathcal L}{\partial \dot q^k}= \frac{1}{\tN}g_{kl}\dot q^l+\mathcal Q_k(\Psi) \,,
\ee
leads to a first-order action of the form
\be \label{Haction}
 S=\int dt \Big( p_k \dot q^k+   \ft i2 \,G_{ab}\left({  \Phi}^{\dagger a} \dot\Phi^b-\dot{\Phi}^{ \dagger a}\Phi^b\right)- \tN H^{\rm tot}\Big) \,,
\ee
with
\bea \label{Htot0}
H^{\rm tot}&=& \frac12  g^{kl} (p_k -  \mathcal Q_k) (p_l -  \mathcal Q_l) -  L_{\Psi^4}\nonumber\\
&=&\ft 12 g^{kl} p_k p_l -  g^{kl}  p_k \mathcal Q_l
+\ft 12 g^{kl}\mathcal Q_k\mathcal Q_l -  L_{\Psi^4},
\eea
where $ g^{kl}$ denotes the inverse of the symmetric quadratic form $g_{kl}$ defining the bosonic kinetic terms. As the  $\mathcal Q_k$'s are bilinear in the fermions, the term 
$\ft 12 g^{kl}\mathcal Q_k\mathcal Q_l $ adds to the original quartic-in-fermions term $ -  L_{\Psi^4}$.

The structure of the Hamiltonian action \eqref{Haction} shows that $\tN$ is a Lagrange multiplier,
associated with the Hamiltonian constraint
\be
H^{\rm tot} =0\,.
\ee
In addition, the explicit computation of $H^{\rm tot} $ shows (as guaranteed by the local
supersymmetry of the original, unreduced supergravity action) that the time component $\Psi_{\hat 0}$
of the gravitino (and its Dirac conjugate $\overline \Psi_{\hat 0}$) appear only linearly in $H^{\rm tot}$.
They are therefore two fermionic Lagrange multipliers, associated with two supersymmetry constraints, say
\be
{\cal S} =0 \; , \; {\overline {\cal S}}=0 \,,
\ee
whose expressions will be given below.

The computation of $H^{\rm tot}$ leads to an expression of the form
\be \label{Htot1}
H^{\rm tot}=H^{(0)}+ H^{(2)}+H^{(4)}+ {\overline\Psi}'_{\hat 0}{\cal S}+ {\overline{\cal S}}{\Psi}'_{\hat 0}\,,
\ee
where the superscripts indicate the polynomial order in the {\it spatial} components, $\Psi_\ha$, or $\overline \Psi_{\hat a}$, of the gravitino, and where we introduced the following shifted time component of the
gravitino:
\be \label{Psi0'}
{\Psi}'_{\hat 0} \equiv {\Psi}_{\hat 0}- \gamma_{\hat 0}\sum_a\gamma^{\hat a}\Psi^{\hat a}\,.
\ee
The terms in Eq. \eqref{Htot1} read as follows.

The purely bosonic part of the Hamiltonian reads
\be \label{H0}
H^{(0)}= G^{ab} \pi_a \pi_b + 2 \sum_{a<b} e^{-2(\beta^{ b}-\beta^{ a})} ({P}_{ab})^2+ \frac12 \sum_a e^{-2\,\beta^a}(P^a)^2\,,
\ee
where $\pi_a$ is the conjugate momentum to $\beta^a$,
$P^a$ is the momentum conjugate to $B_a$, and where ${P}_{ab}$ (with $a<b$) is the following combination
of the conjugate momentum ${p^i}_a $ to  ${n^a}_i $ and of $P^a$,
\be \label{Pab}
{P}_{ab} \equiv \sum_{a<i \leq b} {p^i}_a  {\N^b}_i - B_a P^b\,.
\ee
The part of the Hamiltonian that is quadratic in fermions reads
\be \label{H2}
H^{(2)}= + 2 \sum_{a<b} e^{-(\beta^{ b}-\beta^{ a})} {P}_{ab} J_{ab}(\Psi) - \frac{1}{\sqrt{3}}\sum_a e^{-\beta^a}P^a J_a(\Psi)\,,
\ee
where $ J_{ab}(\Psi)$ and  $J_{a}(\Psi)$ are fermion bilinears (defined in Eqs. \eqref{defJ} below).

The part of the Hamiltonian that is quartic in fermions is given by the following sum
\be \label{H4}
H^{(4)}= \frac12 \sum_{a<b} \left(J_{ab}(\Psi)\right)^2 + \frac16 \sum_{a} \left(J_{a}(\Psi)\right)^2 
- L_{\Psi^4}^{\rm cg}\,,
\ee
where the superscript cg  means that one should replace everywhere in $L_{\Psi^4}$ ${\Psi}_{\hat 0}$
by its ``coset gauge value'', ${\Psi}_{\hat 0}^{\rm cg}$, obtained by setting ${\Psi}'_{\hat 0} $ to zero, 
{\it i.e.}, in view of Eq. \eqref{Psi0'}, by 
\be
{\Psi}_{\hat 0}^{\rm cg}= \gamma_{\hat 0}\sum_a\gamma^{\hat a}\Psi^{\hat a}= \gamma_{\hat 0}\sum_a\Phi^{\hat a}\,.
\ee
 It was found in previous works that this coset gauge has
the property of revealing hidden Kac-Moody structures in the fermionic dynamics.

The fermion bilinears  $J_{ab}(\Psi) $ (with $a<b$) and $J_{a}(\Psi) $ entering both $H^{(2)}$ and $H^{(4)}$ have the
factorized vector-spinor structure found in Ref. \cite{Damour:2009zc} (and generalized in Refs.  \cite{Kleinschmidt:2013eka,Kleinschmidt:2018hdr})
, namely
\bea \label{defJ}
J_{ab}(\Psi) &=& (G_{cd}- 2\alpha^{(ab)}_c \alpha^{(ab)}_d) \Phi^{\dagger\,c}\left(\frac{i \gamma^{ a b}}{2}\right)\Phi^d \,,  \nonumber\\
J_{a}(\Psi) &=& (G_{cd}-2 \alpha^{(a)}_c \alpha^{(a)}_d) \Phi^{\dagger\,c}\left(\frac{3 \gamma^{ a}}{2} \right)\Phi^d \,,
\eea
where $\alpha^{(ab)}_c$ and $\alpha^{(a)}_c$ denote the (covariant) components of the 
linear forms in the $\beta$'s that appear as exponents in several pieces of the Hamiltonian, namely
\bea \label{defalpha}
\alpha^{(ab)}(\beta) &\equiv&\alpha^{(ab)}_c \beta^c \equiv \beta^b-\beta^a \,,\nonumber\\
\alpha^{(a)}(\beta) &\equiv&\alpha^{(a)}_c \beta^c \equiv \beta^a \,.
\eea
For instance, $\alpha^{(a)}_c =\delta^a_c$. The (Kac-Moody) meaning of the linear forms 
$\alpha^{(ab)}(\beta)$, $\alpha^{(a)}(\beta)$ wil be explained in the next section. Note that in
the definitions \eqref{defJ},   $a$ and $b$ are numerical labels (which are not summed over),
 while $c$ and $d$ are vectorial indices in $\beta$ space that are summed over as per the Einstein convention.

Finally, the supersymmetry constraint $\cal S$ has the form
\be\label{Stot}
{\cal S}= {\cal S}^{(1)}  + {\cal S}^{(3)}\,,
\ee
where the linear in fermion part is,
\bea \label{S1}
{\cal S}^{(1)}&=& \sum_a \pi_a \Phi^a - \sum_{a<b}  e^{-(\beta^{ b}-\beta^{ a})} {P}_{ab} \gamma^{ab} (\Phi^b-\Phi^a) \nonumber\\
&-& i\frac{\sqrt{3}}{2} \sum_a e^{-\beta^a}P^a \gamma^a \Phi^a\,,
\eea
while the cubic in fermion part reads
\begin{widetext} 
 \eqal{\label{cS3}
 {\cal S}^{(3)}=&-\ft 12\sum_{{\hat p},{\hat q}} ({\overline\Psi}_{\hat p}\gamma_{\hat q}\Psi_{\hat q}-{\overline\Psi}_{\hat q}\gamma_{\hat q}\Psi_{\hat p})\ \gamma_{\hat  0}\Psi_{\hat p}-\ft 12\sum_{{\hat q},{\hat p}>{\hat q}} ({\overline\Psi}_{\hat p}\gamma_{\hat  0}\Psi_{\hat q}-{\overline\Psi}_{\hat q}\gamma_{\hat  0}\Psi_{\hat p}) \ \gamma_{\hat p}\Psi_{\hat q}\nonumber{}\\
&-\ft {i}{2}  \sum_{{\hat p},{\hat q},{\hat r},{\hat s} }\eta^{{\hat  0} {\hat p}{\hat q}{\hat r}{\hat s}}({\overline\Psi}_{\hat p}\gamma_{\hat q}\Psi_{\hat r})\ \Psi_{\hat s}+\ft i2 \sum_{{\hat p},{\hat q},{\hat s} ,{\hat r}>{\hat s}}\eta^{{\hat  0} {\hat p}{\hat q}{\hat r}{\hat s}}({\overline\Psi}_{\hat p}\Psi_{\hat q})\ \gamma_{\hat s}\Psi_{\hat r}\nonumber{}\\
&+\ft i{4}\sum_{\hat q,\hat r,\hat s,\hat k,{\hat p}>{\hat k}}\eta^{{\hat  0}{\hat p}{\hat q}{\hat r}{\hat s}}({\overline\Psi}_{\hat k}\gamma_{{\hat q}{\hat r}}\Psi_{\hat s}+{\overline\Psi}_{\hat s}\gamma_{{\hat q}{\hat r}}\Psi_{\hat k})\,(\gamma_{\hat k}\Psi_{\hat p}+\gamma_{\hat p}\Psi_{\hat k}) 
+\ft i{4}\sum_{\hat q,\hat r,\hat s,{\hat p}}\eta^{{\hat  0}{\hat p}{\hat q}{\hat r}{\hat s}}({\overline\Psi}_{\hat p}\gamma_{{\hat q}{\hat r}}\Psi_{\hat s}+{\overline\Psi}_{\hat s}\gamma_{{\hat q}{\hat r}}\Psi_{\hat p})\ \gamma_{\hat p}\Psi_{\hat p} 
\,.}
\end{widetext}

%%%%%%
\section{Intermezzo on the hyperbolic Kac-Moody algebra $G_2^{++}$, and its maximally compact subalgebra $K(G_2^{++})$.}

The bosonic part of the Hamiltonian,
\be \label{H0bis}
H^{(0)}= G^{ab} \pi_a \pi_b + 2 \sum_{a<b} e^{-2(\beta^{ b}-\beta^{ a})} ({P}_{ab})^2+ \frac12 \sum_a e^{-2\,\beta^a}(P^a)^2\,,
\ee
 can be viewed (when remembering the constraint $H^{(0)}$=0)
 as describing the dynamics of a 
{\it massless} particle (submitted to the constraint $g_{kl} \dot q^k  \dot q^l=0$),
 with coordinates  $q^k =(\beta^{ \hat a} \,;\, {n^{\hat a}}_{ i}\,;\, B_{\hat a})$ 
 [or, equivalently, $(h_{ij}; A_i)$] moving
in a 14-dimensional curved (Lorentzian-signature) spacetime, with metric $ds^2= g_{kl} d q^k  d q^l$
defined by 
\bea
ds^2&=& G_{ab} d\beta^{ a} d\beta^{ b}+\ft12 \sum_{a<b}e^{+2({\beta}^{  b}-{\beta}^{ a})} \left( dn^{ a}_{\ i}({\cal N}^{-1})^i_{\  b}\right)^2  \nonumber\\
&+& \frac1{2}\sum_a e^{2\beta^{ a}}
\left( dB_{ a} + B_{ b}\, {dn}^{b}_{\ i} ({\cal N}^{-1})^i_{\ a} \right)^2\,.
\eea
In terms of the coordinates $h_{ij}, A_i$, this metric reads
\be
ds^2= \frac1{4} \left( h^{ik} h^{jl} -  h^{ij} h^{kl}\right) d h_{ij} d h_{kl}+ \frac12 h^{ij} dA_i dA_j\,.
\ee
Though the latter spacetime metric admits as 20-dimensional 
symmetry group the semi-direct product of $GL(4)$ 
transformations ($\Lambda^i_j$) with $R^4$ translations ($A_i \mapsto A_i +c_i$), its dynamics
is chaotic, and describes the BKL-type chaos of  general
solutions of the Einstein-Maxwell theory near a cosmological singularity \cite{Damour:2000th}.

The  finite-dimensional model defined by Eq. \eqref{H0} is
 a truncation of an infinite-dimensional model describing the dynamics of
 a massless particle on the coset space(time) $G_2^{++}/K(G_2^{++})$,
 where $G_2^{++}$ is the hyperbolic Kac-Moody group defined by the (untwisted) hyperbolic
 extension of the exceptional Lie group $G_2$, and where $K(G_2^{++})$ denotes the
 maximally compact subgroup of $G_2^{++}$, defined as the fixed point of the Chevalley involution (see below).
 The original motivation for considering such an hyperbolic Kac-Moody coset is the fact that the
 four linear forms
 \bea \label{simpleroots}
  \alpha^{(1)}(\beta)&=& \beta^1, \nonumber\\
  \alpha^{(12)}(\beta) &=& \beta^2-\beta^1, \nonumber\\
  \alpha^{(23)}(\beta) &=& \beta^3-\beta^2, \nonumber\\  
  \alpha^{(34)}(\beta) &=& \beta^4-\beta^3, 
 \eea
 entering the four dominant potential walls (among the Toda-like potentials 
 $e^{-2({\beta}^{  b}-{\beta}^{ a})}$,  $e^{-2 {\beta}^{ a}}$ of the bosonic Hamiltonian Eq. \eqref{H0})
 that determine its chaotic behavior, can be identified with the four simple roots of  $G_2^{++}$.
 Indeed, the four linear forms $\alpha_i(\beta) = \alpha_{i \, a} \beta^a$, $i=1,2,3,4$, with $\alpha_1\equiv\alpha^{(1)}$,
 $\alpha_2 \equiv \alpha^{(12)}$, $\alpha_3 \equiv \alpha^{(23)}$, $\alpha_4 \equiv\alpha^{(34)}$,
 viewed as forms in $\beta$ space, with metric $G_{ab}$ (so that we have the scalar product
 $\langle \alpha_{i }, \alpha_{j }\rangle \equiv\alpha _{ i\,a}G^{ab}\alpha _{j\, b}$)
  have squared lengths equal to
 \be
 \langle \alpha_1,\alpha_1\rangle=\ft23 \;, \;\langle \alpha_i, \alpha_i\rangle=2\; {\rm for}\; i=2,3,4\,.
 \ee
 The associated Cartan matrix (which define $G_2^{++}$) :
\eqal{A_{ij}=2\frac{\langle \alpha_i, \alpha_j\rangle}{\langle \alpha_i, \alpha_i\rangle} \,,
}
is given by
\be \label{Cartan}
(A_{ij})= 
\left(
\begin{array}{cccc}
 2 &-3   &0&0   \\
-1  &2   &-1&0   \\
 0 & -1  & 2&-1\\
 0&0&-1&2  
\end{array}
\right) \,.
\ee
The corresponding Dynkin diagram is represented in  Eq. \eqref{dynkinG2pp}:
\begin{equation} \label{dynkinG2pp}\hspace{10mm}<\hspace{-11.5mm}
\xymatrix{
{ \underset{\alpha^{(1)}}{ \bullet}\!\!\!} \ar@{=}[r]\ar@{-}[r] &{ \!\!\!\underset{\alpha^{(12)}}{ \bullet} \!\!\!}   \ar@{-}[r] &{{\!\!\! \underset{\alpha^{(23)}}{ \bullet}\!\!\! } }\ar@{-}[r]&{ \!\!\!\underset{\alpha^{(34)}}{ \bullet} \!\!\!}
}
\end{equation}
The Chevalley-Serre-Kac presentation is then defined by the four  $sl(2)$ triplets $( e_i, h_i,f_i)$,
$i=1,2,3,4$ (associated with the four simple roots $\alpha_i$), satisfying the standard defining relations of a Kac-Moody algebra associated with $A_{ij}$:
\be
[h_i,h_j] = 0; [e_i,f_j] = \delta_{ij} h_j ;[h_i,e_j] = A_{ij} e_j ; [h_i,f_j] = - A_{ij} f_j
\ee
together with the crucial Serre relations
\be
{\rm ad}(e_i )^{1-A_{ij}} (e_j ) = 0 \,; \, {\rm ad}(f_i )^{1-A_{ij}} (f_j ) = 0 \,.
\ee
Summarizing the present section so far, the linear forms $\alpha^{(ab)}(\beta)$, $\alpha^{(a)}(\beta)$
entering the (Bianchi-I-reduced) bosonic Hamiltonian \eqref{H0} suffice to characterize the hyperbolic
Kac-Moody algebra $G_2^{++}$. Similarly to the decomposition of $E_{10}$ associated
with eleven-dimensional supergravity \cite{Damour:2002cu,Damour:2002et}, one can decompose
the Lie algebra of $G_2^{++}$ with respect to the $gl(4)$ subalgebra  defined by the 
gravity-related roots $\alpha^{(ab)}(\beta)$ (together with the Cartan element $h_1$) \cite{Mizoguchi:2005zf}.
 One adds to this level-0 subalgebra ($K^a_{\ b}$) the 
level-1 generators $E^a$ and $F_a$ associated with the electric-related roots 
$+\alpha^{(a)}(\beta)$ and $- \alpha^{(a)}(\beta)$. The rest of the algebra is then defined
by taking commutators, starting with the level-2 defined by $ E^{[ab]} \equiv [E^a,E^b]$, 
the level 3 defined by $ E^{[abc]} \equiv [E^a,E^{[bc]}]$, etc.
It it then checked \cite{Mizoguchi:2005zf} that the bosonic dynamics defined by Eq. \eqref{H0} is equal
to the reduction of the  infinite-dimensional  coset dynamics on $G_2^{++}/K(G_2^{++})$
obtained by setting to zero the momenta corresponding to all the positive roots of
levels $\ell \geq 2$ (similarly to the truncation of $E_{10}/K(E_{10})$ beyond level 1 
\cite{Damour:2002cu,Damour:2002et}). 

The general conjecture made in Refs. \cite{Damour:2002cu,Damour:2002et} is that
there is a gravity-coset correspondence under which the dynamics of any supergravity theory
would be equivalent to a corresponding hyperbolic Kac-Moody coset dynamics, having the same
asymptotic cosmological billiard. 
The purpose of the present work is to obtain new evidence for such a correspondence
by focussing on the fermionic sector of 5-D supergravity, and particularly on the terms
quartic in fermions, which have been neglected in most of the previous investigations of the
gravity-coset conjecture. In this respect, we need to consider in detail the coset analog
of the R-symmetry, {\it i.e.}, the symmetry group under which the coset fermions are
conjectured to rotate. This group is supposed (in each coset model) to be the 
maximally compact subgroup of the considered Kac-Moody group. In the case of 5-D
supergravity, this is  $K(G_2^{++})$, whose Lie-algebra is defined as the fixed point of 
the Lie algebra of $G_2^{++}$ under the Chevalley involution. The Chevalley involution $\theta$
is defined by its action on the Chevalley-Serre-Kac basis: 
\be
\theta(h_i) = - h_i, \theta(e_i) = - f_i, \theta(f_i) = - e_i \,.
\ee
The $\theta$-fixed subalgebra $K(G_2^{++})$ of  $G_2^{++}$ is then generated by the four
Lie-algebra elements
\be
x_i \equiv e_i - f_i\,.
\ee
Previous work on supergravity in D=11  has shown that the gravitino field belonged to a finite-dimensional representation of the (infinite-dimensional) Lie algebra $K(E_{10})$ \cite{Damour:2005zs,de Buyl:2005mt,Damour:2006xu}. Analog results
were found for other supergravity theories \cite{Damour:2002fz}. In our present context, we therefore expect that
the 5-D gravitino $\Psi_a$ will belong to a finite-dimensional representation of $K(G_2^{++})$.
The main results of the present work will indeed be to show not only that this is true, but to
further show that the $O(\Psi^4)$ term in the (quantum) Hamiltonian is invariant under the $K(G_2^{++})$
rotations defining the representation of the quantized gravitino. In order to investigate technically this issue
we will need to characterize the conditions defining a representation of $K(G_2^{++})$.

A linear representation of $K(G_2^{++})$ is characterized by a vector space on which acts four
linear operators $J_i$ satisfying the same defining relations as the four 
abstract Lie-algebra elements $x_i=e_i-f_i$ defined above. As each operator $J_i$ (corresponding to $x_i$)
is associated to the specific simple root $\alpha_i$, it will be convenient to label the linear operators $J_i$
by the same label as the associated simple root of $G_2^{++}$, as listed in Eq. \eqref{simpleroots}.
Therefore, we will denote them simply as $J_1$, $J_{12}$, $J_{23}$ and $J_{34}$, respectively
associated with $\alpha^{(1)}(\beta)$, 
  $\alpha^{(12)}(\beta)$,  $ \alpha^{(23)}(\beta)$ and $\alpha^{(34)}(\beta)$.

The set of defining (Serre-Berman) relations that the four operators $J_1$, $J_{12}$, $J_{23}$, $J_{34}$ must satisfy is
\cite{Berman,Koehl}
\eqal{\label{berman}
&{\rm ad}^4(J_1)J_{12}-10\,{\rm ad}^2(J_1)J_{12}+9\,J_{12}=0 \,,\\
&{\rm ad}^2(J_{12})J_{1}-J_1=0\,,\\
&{\rm ad}^2(J_{12})J_{23}-J_{23}=0\,,\\
&{\rm ad}^2(J_{23})J_{12}-J_{12}=0\,,\\
&{\rm ad}^2(J_{23})J_{34}-J_{34}=0\,,\\
&{\rm ad}^2(J_{34})J_{23}-J_{23}=0\,,\\
&[J_1,J_{23}]=[J_1,J_{34}]=[J_{12},J_{34}]=0 \,.}
Note that in the present work we will be dealing with hermitianlike rotation operators $J_i$,
instead of the anti-hermitian ones $x_i$ used in mathematical contexts. In other words, a
$J_i$ rotation will be of the type $\exp ( \sqrt{-1} \theta_i J_i)$, instead of $\exp( \theta_i {\rm ad} x_i)$.

Several different representations of $K(G_2^{++})$ play a role in our present 5-D supergravity context.
First, there are representations associated with classical ({\it i.e.} grassmanian-valued) fermions, of
spin $\frac12$ and $\frac32$. Second, there are representations of $K(G_2^{++})$ associated with
quantum fermions. Let us describe now the representations of classical spinors, of spin $\frac12$ and $\frac32$.

Note first that the relations involving only the $J_{ab}$'s among Eqs. \eqref{berman} express the fact
that the $J_{ab}$'s are usual $SO(4)$ rotation operators. The four (complex) components
of a spin $\frac12$ Dirac field $\Psi$ define a representation space for the  $J_{ab}$'s if we
define their action in the usual ${\rm Spin}(4)$ way, namely
\be
J_{ab}^s= i\frac{ \gamma^{ab}}{2}\; ; \; 1\leq a < b \leq4 \,.
\ee
We defined here not only the three simple-root generators  $J_{12}$, $J_{23}$, $J_{34}$
(associated with the symmetry-wall simple roots), but also
the three others needed to describe the rotations in all the two-planes $ab$ of $R^4$. [The factor $i$
is needed because we are working with hermitianlike operators.]
It is then easy to check that if we tentatively define the generators $J_a$ associated with the
electric roots $\alpha^{(a)}(\beta)=\beta^a$ as
\be \label{Jas}
J_{a}^s= \frac{ C^s \gamma^{a}}{2}\; ; \; 1\leq a \leq4 \,,
\ee
the defining relations Eqs. \eqref{berman} will be satisfied if the factor $C^s$ is equal to
\be
C^s= \pm 1 \; ; \; {\rm or} \; C^s= \pm 3 \; ({\rm for \ a \ spinor \ representation}) \,.
\ee
Indeed, the first defining relation can actually be factorized as
\be
\left( {\rm ad}^2(J_1) - 3^2 \right)  \left( {\rm ad}^2(J_1) - 1^2 \right) J_{12} = 0 \,.
\ee
We can then define two types of 4-dimensional spinor representations of $K(G_2^{++})$ (with $C^s= \pm 1$,
or $C^s= \pm 3$).

Let us now consider the possible vector-spinor representations of $K(G_2^{++})$, {\it i.e.}, matrices
$J_{ab}$, $J_a$, acting on the sixteen components of a gravitinolike object $\Phi^a = \gamma^a \Psi^a$.
[Here the vector index $a=1,2,3,4$ and the hidden Dirac-spinor index takes also four values.]
Here we consider the actions on $\Phi^a$ rather than on $\Psi^a$ because it was found in 
Ref. \cite{Damour:2009zc} that this reveals a hidden factorized structure for the vector-spinor
representations associated with $K(E_{10})$ and $K(AE_{3})$. We found that such a factorized
structure also holds for $K(G_2^{++})$, in spite of the fact that $G_2^{++}$ is not simply laced
(remember that the simple root $\alpha_1$ has length-squared $\frac23$). More precisely,
we define, for any one of the simple roots (and more generally
for any of the basic gravitational or electric roots entering the levels 0 and 1), the action of $J_{\alpha}$ on
a vector spinor $\Phi^{aA}$ as
\be \label{Jvsalpha}
\left( J^{vs}_{\alpha} \cdot \Phi \right)^{aA} \equiv (\delta^a_b - 2 \alpha^a \alpha_b) \left(J^s_{\alpha}\right)^A_{\ B} \Phi^{bB} \,, \
\ee
where $\alpha^a \equiv G^{ac} \alpha_c$, and where $J^s_{\alpha}$ is the above-defined action of 
$J_{\alpha}$ in the (4-dimensional) spinor representation.
Here, for clarity, we have explicitly indicated the (usually implicit) spinor indices $A,B$. We then found
(in agreement with Ref. \cite{Kleinschmidt:2018hdr}) that the
vector-spinor matrices $ (J^{vs}_{\alpha})^ {aA}_{\ \ bB}$ satisfy the defining relations \eqref{berman} if,
and only if, the factor $C^s$ entering the electric operator \eqref{Jas} is taken to be
\be
C^{s}= \pm 3 \; ({\rm for \ a \ vector-spinor \ representation}) \,.
\ee
The value $C^s=\pm 1$ fails to define a vectorspinor representation of $K(G_2^{++})$ when inserted in
Eq. \eqref{Jvsalpha}. Note that this is precisely the value $C^s=3$ that appeared in the supergravity-derived
bilinear $J_a$, Eq. \eqref{defJ}. We shall explain below, after quantizing the supergravity dynamics, the
meaning of the quantum avatars of the bilinears in Eq. \eqref{defJ} as generators of a $2^{16}$-dimensional
representation of $K(G_2^{++})$, in which lives the quantum state of our cosmological model.

%%%%%%
\section{Quantization}

The classical Hamiltonian action of our supersymmetric cosmological model has the form
(with ${\N^a}_i \equiv \delta^a_i+ n^a_{\ i}$)
\bea \label{Haction2}
 S&=&\int dt \Big( \pi_a \dot \beta^a+ {p^i}_a  \dot{n^a}_i + P^a \dot B_a+ \ft i2 \,G_{ab}\left({  \Phi}^{\dagger a} \dot\Phi^b-\dot{\Phi}^{ \dagger a}\Phi^b\right) \nonumber\\
 && - \tN \left( H^{\rm cg}+ {\overline\Psi}'_{\hat 0}{\cal S}+ {\overline{\cal S}}{\Psi}'_{\hat 0}\right)\Big). \nonumber\\
\eea
Here we use Einstein's summation convention. The indices $ a, i$ of the (strictly) upper triangular matrix 
$ n^a_{\ i}$, and therefore those of its canonical conjugate ${p^i}_a$ are restricted to the range $a<i$.
The contribution $H^{\rm cg}$ to the total Hamiltonian has the structure
\be
H^{\rm cg}=  H^{(0)}+ H^{(2)}+H^{(4)}\,,
\ee
where the explicit values of the terms $H^{(0)}, H^{(2)}, H^{(4)}$ were given above. The action \eqref{Haction2} features three independent Lagrange multipliers: $\tN$, $\tN {\overline\Psi}'_{\hat 0}$,
and $\tN {\Psi}'_{\hat 0}$, where ${\Psi}'_{\hat 0}$ is the shifted value of  ${\Psi}_{\hat 0}$,
defined in Eq. \eqref{Psi0'}. 
These Lagrange multipliers reflect the presence of three local-in-time
gauge symmetries: (i) invariance under reparametrization of the time variable; and (ii) the two
local-in-time supersymmetries $\epsilon_0(t), {\overline \epsilon_0}(t)$. These gauge symmetries
allow one to choose at will the values of the Lagrange multipliers $\tN $ and  $\tN {\Psi}'_{\hat 0}$.
It is convenient to choose the coset gauge where $\tN=1$ and ${\Psi}'_{\hat 0}=0$.

The  action \eqref{Haction2} defines a constrained dynamics, with first-class constraints
\be \label{constraints1}
H^{\rm cg} \approx 0 \, ; \, {\cal S}_A \approx 0 \, ; \, {\overline{\cal S}}_A \approx 0\,,
\ee
where we explicitly indicated the spinor index $A$, which takes four values.

The classical ({\it i.e.}, Grassmanian) consistency of supergravity implies that the constraints 
\eqref{constraints1} close under the Poisson(-Dirac) brackets defined by the kinetic terms
\bea
&&\{\beta^a, \pi_b \}_P= \delta^a_b\,;\, \{n^a_{\ i}, p^j_{\ b} \}_P =  \delta^a_b  \delta^j_i \, ; \nonumber\\
&&\{B_a, P^b \}_P=\delta^a_b\,;\, \{\Phi^{a A}, {\Phi^{\dagger}}^{bB} \}_P= \frac1{i} G^{ab} \delta^{AB}.
\eea
A crucial classical identity (which we checked to hold) is the fact that the Poisson brackets 
of the supersymmetry constraints close as follows:
\bea \label{cidentities}
\{ {\cal S}_A , {\cal S}_B \}_P &=&0 \, ; \nonumber\\
\{ {\cal S}^{\dagger}_A , {\cal S}^{\dagger}_B \}_P &=&0 \, ; \nonumber\\
\{ {\cal S}_A , {\cal S}^{\dagger}_B \}_P &=& {L(\Phi)^{\dagger}}^C_{AB} {\cal S}_C -  {L(\Phi)}^C_{AB} {\cal S}^{\dagger}_C + \frac1i \delta_{AB} H^{\rm cg}\,, \nonumber\\
\eea
where ${L(\Phi)}^C_{AB} $ is linear in $\Phi$ (and does not contain $\Phi^{\dagger}$, nor any of the
bosonic dynamical variables).

We quantize the constrained dynamics defined by the  action \eqref{Haction2} \`a la Dirac, {\it i.e.} 
by: (i) replacing Poisson-Dirac brackets by appropriate (anti-)commutators; (ii) verifying that this allows one to construct operators providing a deformed version of the classical algebra of constraints; and (iii) imposing the quantum constraints $\widehat{\mathcal C}= (\widehat H, \widehat {\cal S}_A, \widehat {\cal S}^{\dagger}_A)$ as conditions restricting physical states $\vert{{\bm \Psi}}\rangle$: $\widehat{\mathcal C} \vert{{\bm \Psi}}\rangle = 0$.

For the bosonic degrees of freedom we adopt a Schr\"odinger picture. The wave function of the Universe is seen as a function of the fourteen configuration-space variables $\beta^a$, $n^a_{\ i}$ and $B_a$. 
The corresponding basic conjugate quantum momenta operators are represented as 
\begin{eqnarray} \label{qmomenta}
\widehat\pi_a &= &\frac{\hbar}{i} \, \partial_{\beta^a} \,,\nonumber \\
\widehat p^i_{ \ a} &= &\frac{\hbar}{i} \, \partial_{n^a_{\ i}} \,, \nonumber\\
\widehat P^a &= &\frac{\hbar}{i} \, \partial_{B_a}\,.
\end{eqnarray}
In the following, we shall often set $\hbar = 1$.
The  momentum-like combination $P_{ab}$, Eq. (\ref{Pab}), associated with the symmetry walls $\alpha^{(ab)}$ are then defined as
\bea \label{Pabq}
\widehat{P}_{ab} &\equiv& \sum_{a<i \leq b} {\widehat p^i}_a  {\N^b}_i - B_a \widehat P^b \nonumber\\
&=& \sum_{a<i \leq b}  {\N^b}_i  {\widehat p^i}_a -  \widehat P^b B_a\,.
\eea
As indicated, there are no ordering ambiguities in defining $\widehat{P}_{ab}$ because they are
defined only for $a<b$. These operators satisfy an algebra which coincides (modulo a factor $i \hbar$)
with the classical one. For instance, we have
\be
\{ P_{12}, P_{23} \}_P = P_{13} \, ; \, {\rm and} \;  [\widehat P_{12}, \widehat P_{23} ] = i \hbar \widehat P_{13}\,.
\ee
Similarly
\bea
\{ P_{12}, P^1 \}_P &=& - P^2 \, ; \, {\rm and} \;  [\widehat P_{12}, \widehat P^1 ] =- i \hbar \widehat P^2 \nonumber \\
\{ P_{12}, P^2 \}_P &=& 0 \, ; \, {\rm and} \;  [\widehat P_{12}, \widehat P^2 ] = 0\,.
\eea

The fermionic operators have to obey anticommutations relations dictated by their kinetic term.
These anticommutation relations take an especially simple form when using the objects $\Phi^a$
and ${\Phi^{\dagger}}^a$ (rather than $\Psi^a$ and $\overline \Psi^a$), namely
\bea \label{CRPhi}
\{ \widehat \Phi^{aA} , {\widehat \Phi}^{ bB} \}&=&0\,,\, \{ \widehat \Phi^{\dagger aA} , {\widehat \Phi}^{\dagger bB} \}=0 \, ,\nonumber\\
\{ \widehat \Phi^{aA} , {\widehat \Phi}^{\dagger bB} \} &=& \hbar G^{ab} \, \delta^{AB} \,,
\eea
where, now, the curly brackets (without a $P$ subscript) denote an anticommutator.

%This shows that the thirty-two quantum fermionic operators $\widehat \Phi^{Aa}$, ${\widehat \Phi}^{\dagger aA}$  satisfy a Clifford algebra in a thirty-two-dimensional space. 
When decomposing the hermitian-conjugated quantum fermionic operators $\widehat \Phi^{Aa}$, 
${\widehat \Phi}^{\dagger aA}$ into their (formally) hermitian parts, 
$\widehat \phi^{Aa}_1\equiv \frac12 (\widehat \Phi^{Aa}+ {\widehat \Phi}^{\dagger aA})$,  
$\widehat \phi^{Aa}_2\equiv\frac1{2 i} (\widehat \Phi^{Aa}- {\widehat \Phi}^{\dagger aA})$,
the  thirty-two fermionic operators, $\widehat \phi^{Aa}_i$, $i=1,2$, are found to satisfy a Clifford algebra
 in a {\it real} thirty-two-dimensional space endowed with the quadratic form 
 $\frac12 \hbar \,\delta_{ij}  G^{ab} \, \delta^{AB}$, which has signature $24^+, 8^-$.
 Thus the gravitino operators can be represented by $2^{16}\times 2^{16} = 65536\times65536$ 
 Dirac matrices\footnote{In view of the signature $24^+, 8^-$, these matrices can be chosen to be real.} 
 and the wave function of the Universe can be viewed as
  a 65536-dimensional spinor of Spin(24,8), depending on  the fourteen configuration-space variables 
  $\beta^a$, $n^a_{\ i}$ and $B_a$:
 ${\bm \Psi} = {{\Psi}}_{\sigma} (\beta^a , \varphi^b)$, with $\sigma = 1,\ldots ,65536$.

 %%%%%%
\section{Quantum constraints and their consistency} \label{qconstraints}
 
A crucial issue  in the quantization of our system is to promote the classical constraints \eqref{constraints1}
  into corresponding quantum operators, say  $\widehat {\cal S}_A$, 
 $\widehat {\overline {\cal S}}_A$, and $\widehat H$, so as to impose them, \`a la Dirac, on the state $\vert{\bm\Psi}\rangle$: 
\begin{equation}
\label{qconstraints}
\widehat {\cal S}_A \, \vert{\bm\Psi}\rangle = 0 , \; \widehat {\overline {\cal S}}_A  \, \vert{\bm\Psi}\rangle = 0 , \;\widehat H \, \vert{\bm\Psi}\rangle = 0 \, .
\end{equation}
However, such a quantization scheme will be consistent only
 if we can define an ordering such that the quantum constraints
operators $\widehat {\cal S}_A$, 
 $\widehat {\overline {\cal S}}_A$, and $\widehat H$ do close on themselves by satisfying a quantum version of the classical identities \eqref{cidentities}. Let us indicate how we succeeded in defining such an ordering and then in proving its
 quantum consistency.
 
 The structure of the classical supersymmetry constraint is, sketchily,
 \be
 {\cal S}\sim \pi \Phi + \sum_\alpha e^{- \alpha(\beta)} P_{\alpha} \Phi + \Phi^{\dagger} \Phi \Phi\,.
 \ee
 There are no ordering ambiguities in the dependence of $ {\cal S}$ on bosonic variables because: (i) 
 the bosonic variables commute with the fermionic ones; (ii) the wall forms $\alpha(\beta)$ commute with the
 momentalike variables $P_\alpha$; and (iii) we have seen that the $P_\alpha$'s have no internal 
 ordering ambiguities. Finally, the only ordering ambiguity in the definition of $ \widehat {\cal S}$ is
 contained in the last, cubic-in-fermions term $ \Phi^{\dagger} \Phi \Phi$. The ordering of the latter term
 is, however, uniquely fixed by the natural requirement of respecting the symmetry between the $\Phi$'s
 and the  $\Phi^{\dagger}$'s that is present in the basic quantization conditions \eqref{CRPhi}.

 Starting from the classical (Grassmannian) expression of the cubic contribution, 
 \be
 {\cal S}^{(3)}_A= + \sigma_A^{ {\bar {\cal C}} [{\cal A} {\cal B}]} \overline \Psi_{\bar {\cal C}} \Psi_{\cal A} \Psi_{\cal B}\,,
 \ee
 we define its quantum version by
 \be \label{qS3}
\widehat {\cal S}^{(3)}_A \equiv -  \sigma_A^{ {\bar {\cal C}} [{\cal A} {\cal B}]}   \widehat \Psi_{\cal A}  \widehat {\overline \Psi}_{\bar {\cal C}} \widehat \Psi_{\cal B}\,.
 \ee
 Here, the calligraphic indices ${\cal A}, {\cal B}, {\bar {\cal C}}$ denote combined vector-spinor indices
 $aA, bB, cC$ (with an additional bar on the indices pertaining to a ${\overline \Psi}$), and we use the
 Einstein summation convention on these indices. The coefficients 
 $ \sigma_A^{ {\bar {\cal C}} [{\cal A} {\cal B}]}$ are numerical factors (involving products of gamma matrices)
that are defined so as to be antisymmetric in ${\cal A} {\cal B}$.
When reexpressing $\widehat {\cal S}^{(3)}_A $ in terms of  the $\Phi^{\cal A}\equiv\Phi^{aA}$'s and
$\Phi^{\dagger {\cal A}}\equiv \Phi^{\dagger a A}$'s we have
\be \label{qS3Phi}
\widehat {\cal S}^{(3)}_A=  c^A_{ {\cal A} {\cal B}  {\cal C}} \widehat \Phi^{\cal A}  \widehat {\Phi}^{\dagger  {\cal B}} \widehat \Phi^{\cal C}\,,
\ee
with corresponding numerical coefficients  $  c^A_{  {\cal A} {\cal B}  {\cal C}}= -  c^A_{  {\cal C} {\cal B}  {\cal A}}$.
We henceforth use such an ordering\footnote{Actually,  any other
 ordering will lead to the same final physical results because we have shown that any ordering of the cubic terms 
 can be absorbed in a linear
 shift of the $\pi$'s of the type $\pi_a \mapsto \pi'_a=\pi_a + i \delta \pi_a$, where $\delta \pi_a$ are some
 real numbers.}. 

 We have checked that this ordering enforces a symmetry under which $\Phi^{a}$ and ${\widetilde \Phi}^{ a} \equiv \Phi^{\dagger a}$ are swapped:  $\Phi \leftrightarrow \Phi^{\dagger}$.
This is most easily seen by using a representation where $\gamma^0, \gamma^1, \gamma^2, \gamma^3$ are real while $\gamma^4=- i \gamma^0 \gamma^1 \gamma^2 \gamma^3$ is purely imaginary. 
In such a representation the numerical coefficients $  c^A_{  {\cal A} {\cal B}  {\cal C}}$ entering
Eq. \eqref{qS3Phi} are found to be purely imaginary. 
This ensures  that 
\bea \label{updown1}
\left(  c^A_{ {\cal A} {\cal B}  {\cal C}} \widehat \Phi^{\cal A}  \widehat {\Phi}^{\dagger  {\cal B}} \widehat \Phi^{\cal C}\right)^{\dagger} &=&  (c^{A }_{ {\cal A} {\cal B}  {\cal C} })^* \widehat \Phi^{\dagger \cal C}  \widehat {\Phi}^{ {\cal B}} \widehat \Phi^{\dagger \cal A} \\ \nonumber
= -   c^A_{  {\cal A} {\cal B}  {\cal C} }\widehat \Phi^{\dagger \cal C}  \widehat {\Phi}^{ {\cal B}} \widehat \Phi^{\dagger \cal A} &=& +  c^A_{ {\cal A} {\cal B}  {\cal C}} \widehat \Phi^{\dagger \cal A}  \widehat {\Phi}^{ {\cal B}} \widehat \Phi^{\dagger \cal C}\, ,
\eea
where we used $ ( c^{A }_{  {\cal A} {\cal B}  {\cal C} })^*
=-   c^A_{  {\cal A} {\cal B}  {\cal C} }$ together with the antisymmetry  $  c^A_{  {\cal A} {\cal B}  {\cal C}}= -  c^A_{  {\cal C} {\cal B}  {\cal A}}$ and a relabelling of indices, $ {\cal C} \leftrightarrow  {\cal A}$
(which are summed over).

Defining  $ {\cal S}_A$ in the way just explained, we have shown that the following quantum versions of the classical
identities \eqref{cidentities} hold. First,
\bea \label{qidentities1}
\{ \widehat {\cal S}_A , \widehat {\cal S}_B \} &=&0 \, , \nonumber\\
\{ \widehat {\cal S}^{\dagger}_A , \widehat {\cal S}^{\dagger}_B \} &=&0 \, ,
\eea
and, second,
\bea \label{qidentity2}
\{ \widehat {\cal S}_A ,  \widehat{\cal S}^{\dagger}_B \} &=& \frac{i \hbar}{2} [L(\widehat \Phi)^{\dagger  C}_{AB}, \widehat {\cal S}_C ]- \frac{i \hbar}{2} [ {L(\widehat\Phi)}^C_{AB}, \widehat{\cal S}^{\dagger}_C ] \nonumber\\
&&+ \hbar \delta_{AB} \widehat H_0\,, 
\eea
where ${L(\widehat \Phi)}^C_{AB} = L^{C }_{AB  {\cal A}} \widehat \Phi^{\cal A}$ is the {\it same} linear 
form in $\widehat \Phi$ that entered the classical identity \eqref{cidentities}. [The $ L^{C }_{AB  {\cal A}} $
being purely numerical  coefficients made of gamma matrices.] Note the presence of quantum anticommutators
($\{,\}$) on the left-hand side, and the presence of quantum commutators ($[,]$) on the right-hand side. The 
quantum operator $\widehat H_0$ appearing on the last right-hand side is a (formally) hermitian operator
( $\widehat H_0 = \widehat H_0^{\dagger}$), which is a quantum version of the classical Hamiltonian $H^{\rm cg}$.
It has the structure
\be
\widehat  H_0=\widehat  H^{(0)}+ \widehat  H^{(2)}+\widehat H^{(4)}_0
\ee
where the bosonic part reads:
\be \label{qH0}
\widehat H^{(0)}= G^{ab} \widehat\pi_a \widehat\pi_b + 2 \sum_{a<b} e^{-2(\beta^{ b}-\beta^{ a})} (\widehat {P}_{ab})^2+ \frac12 \sum_a e^{-2\,\beta^a}(\widehat P^a)^2\,,
\ee
the part quadratic in fermions reads:
\be \label{qH2}
\widehat H^{(2)}= + 2 \sum_{a<b} e^{-(\beta^{ b}-\beta^{ a})}\widehat {P}_{ab} \widehat J_{ab}(\Psi) - \frac{1}{\sqrt{3}}\sum_a e^{-\beta^a}\widehat P^a \widehat J_a(\Psi)\,,
\ee
where the quantum bilinears $\widehat J_\alpha$ are the quantum avatars of Eqs \eqref{defJ}, namely
\bea \label{defJq}
\widehat J_{ab}(\Psi) &=& (G_{cd}- 2\alpha^{(ab)}_c \alpha^{(ab)}_d) \widehat\Phi^{\dagger\,c}\left(\frac{i \gamma^{ a b}}{2}\right)\widehat\Phi^d \,,  \nonumber\\
\widehat J_{a}(\Psi) &=& (G_{cd}- 2\alpha^{(a)}_c \alpha^{(a)}_d) \widehat\Phi^{\dagger\,c}\left(\frac{3 \gamma^{ a}}{2} \right)\widehat\Phi^d \,,
\eea
and where the quartic-in-fermions part $\widehat H^{(4)}_0$ is a uniquely-defined (hermitian) ordered version of the classical expression \eqref{H4}. There are no ordering ambiguities in the definition \eqref{defJq} of the $\widehat J_\alpha$'s (because
the  matrices $\frac{i \gamma^{ a b}}{2}$ and $\frac{3 \gamma^{ a}}{2}$ are traceless). Actually,
one can also check that the only ordering ambiguity in a hermitian-ordered version of \eqref{H4} lies in
a double Wick contraction, corresponding to an additive $c$-number ambiguity. Anyway, what is important at this stage is
that the existence of the last  identity, Eq. \eqref{qidentity2}, uniquely defines $\widehat  H_0$ and, in particular, 
$\widehat H^{(4)}_0$. If we define an empty state $|0\rangle_-$ as being annihilated by all the $\widehat \Phi$'s,
\be
\Phi^{aA} |0\rangle_- =0,
\ee
we found that
\be \label{VEV0}
\widehat H^{(4)}_0 |0\rangle_- = c_0 \,\hbar^2 \,|0\rangle_- \,,
\ee
where
\be
c_0 = - \frac{743}{24}  \,,
\ee
which characterizes the $c$-number ordering ambiguity in the quantization path leading from the classical $H^{(4)}$ 
to $\widehat  H_0$.

The identity Eq. \eqref{qidentity2} has the advantage of featuring only manifestly hermitian building blocks.
However, it is, by itself, not of the form needed for proving the consistency of our Dirac quantization scheme. Indeed, the
quantum constraints, Eqs.\eqref{qconstraints}, which are of the form $ \widehat {\cal C}_i {\bm \Psi} =0$,  will close on themselves
 only if all the (anti-)commutators between the quantum constraint operators $\widehat {\cal C}_i $ close on 
 this same set of constraints in the following way
 \be
[ \widehat {\cal C}_i , \widehat {\cal C}_j]_\pm = \sum_k \widehat L_{ij}^k \,\widehat {\cal C}_k \,,
 \ee
 with the constraint operators $\widehat {\cal C}_k$ appearing {\it on the right} of the coefficient operators 
 $\widehat L_{ij}^k$. This is not the case for the identity \eqref{qidentity2}, which contains commutators on
 the right-hand side. However, we have shown that the difference between the {\it anticommutator} of 
$ L(\widehat \Phi)^{\dagger  C}_{AB}$ with  $ \widehat {\cal S}_C $, 
and  the one of ${L(\widehat\Phi)}^C_{AB}$ with $\widehat{\cal S}^{\dagger}_C$, is such that it leads
to an identity of the required form, modulo a redefinition of the quantum Hamiltonian $\widehat H$ entering the
last term. More precisely, we found that Eq. \eqref{qidentity2} implies the identity
\bea \label{qidentity3}
\{ \widehat {\cal S}_A ,  \widehat{\cal S}^{\dagger}_B \} &=& i \hbar L(\widehat \Phi)^{\dagger  C}_{AB} \widehat {\cal S}_C - i \hbar {L(\widehat\Phi)}^C_{AB} \widehat{\cal S}^{\dagger}_C  \nonumber\\
&&+ \hbar \,\delta_{AB} \widehat H_1\,, 
\eea
where the new Hamiltonian $\widehat H_1$ reads
\be
\widehat H_1=\widehat H_0 -2\,i\,\varpi^a\widehat \pi_a\,.
\ee
Here the real vector $\varpi^a$ (living in $\beta$ space, or Cartan space) has the following components ($a=1,2,3,4$)
\be
\varpi^a=\frac 14\{1,2,3,4\}
\ee
or, in covariant form ({\it i.e.}, in root space)
\be
\varpi_a \equiv G_{ab}\varpi^a =- \frac 14\{9,8,7,6\}\,.
\ee
The Hamiltonian $\widehat H_0$ which appeared in the identity Eq. \eqref{qidentity2} was (formally) hermitian, while 
the shifted Hamiltonian  $\widehat H_1$ entering the new identity Eq. \eqref{qidentity3} is formally non-hermitian.
A similar situation arose in our previous work \cite{Damour:2014cba}. Like in the latter case, a simple redefinition of the wavefunction 
of the universe allows us to work again with a formally hermitian Hamiltonian. Indeed, if we writes the quantum-state 
wave function $\langle \beta^a , \varphi^b; \sigma | {\bm \Psi}\rangle = {{\Psi}}_{\sigma} (\beta^a , \varphi^b)$ as
\be
 {{\Psi}}_{\sigma} (\beta^a , \varphi^b) = e^{-\varpi_a\beta^a} \, {{\Psi}}_{\sigma}' (\beta^a , \varphi^b)\,,
\ee
the terms involving the differential operator  $\widehat \pi_a = \frac{\hbar}{i} \partial_{\beta^a}$ in 
\bea
\widehat H_1 &=& G^{ab} \widehat \pi_a \widehat \pi_b -2\,i\,\varpi^a\widehat \pi_a + \ldots \nonumber\\
&=& - \hbar^2 G^{ab} \partial_{\beta^a} \partial_{\beta^b} - 2\hbar \varpi^a  \partial_{\beta^a} + \ldots \nonumber\\
&=& - \hbar^2 G^{ab} \left(\partial_{\beta^a} + \varpi_a \right) \left(\partial_{\beta^b} + \varpi_b \right)+ \hbar^2 \varpi^2 + \ldots \nonumber\\
\eea
where we defined
\be
\varpi^2 \equiv G^{ab} \varpi_a \varpi_b = - \frac{70}{16}\,,
\ee
take the following form when reexpressed through their action on ${{\Psi}}_{\sigma}' (\beta^a , \varphi^b)$:
\be
\widehat H_1  {{\Psi}}_{\sigma} (\beta^a , \varphi^b)=  e^{-\varpi_a\beta^a} \widehat H_1^{\prime} {{\Psi}}_{\sigma}' (\beta^a , \varphi^b) \,,
\ee
where
\bea
\widehat H_1^{\prime} &=&  - \hbar^2 G^{ab} \partial_{\beta^a}\partial_{\beta^b} + \hbar^2 \varpi^2 + \ldots \nonumber\\
&=&  G^{ab} \widehat \pi^{\prime}_a \widehat \pi^{\prime}_b + \hbar^2 \varpi^2  + \ldots 
\eea
In the last expression the notation $\widehat \pi^{\prime}_a$ denotes the differential operator $\frac{\hbar}{i} \partial_{\beta^a}$ when acting on the primed wavefunction.

Finally, $\widehat H_1^{\prime}$ can be written as
\be
\widehat  H_1^{\prime}=\widehat  H^{\prime (0)}+ \widehat  H^{(2)}+\widehat H^{\prime (4)}_1 \,.
\ee
Here 
\be \label{qHprime0}
\widehat H^{\prime (0)}= G^{ab} \widehat\pi^{\prime}_a \widehat\pi^{\prime}_b + 2 \sum_{a<b} e^{-2(\beta^{ b}-\beta^{ a})} (\widehat {P}_{ab})^2+ \frac12 \sum_a e^{-2\,\beta^a}(\widehat P^a)^2\,,
\ee
 $\widehat  H^{(2)}$ is given by the same expression \eqref{qH2} as above, and  the last contribution 
is given by
\be
\widehat H^{\prime (4)}_1 = \widehat H^{(4)}_0 + \hbar^2 \varpi^2= \widehat H^{(4)}_0  - \frac{70}{16} \,\hbar^2 \,.
\ee
In view of our previous result \eqref{VEV0}, we conclude that the vacuum value of the new Hamiltonian $\widehat H^{(4)}_1$ is equal to
\be \label{VEV1}
\widehat H^{\prime (4)}_1  |0\rangle_- = c_1 \hbar^2 \,|0\rangle_-  \,,
\ee
where
\be
c_1= c_0 + \varpi^2=  - \frac{743}{24}  - \frac{70}{16}=  - \frac{106}{3}\,.
\ee

%%%%%%
\section{Kac-Moody structure ($G_2^{++}$, $K(G_2^{++})$) of the quantum supergravity dynamics
} \label{KM}

\subsection{Summary of the quantum supergravity dynamics}

Summarizing the results obtained so far, the quantum supergravity dynamics of our five-dimensional cosmological model
is described by a $2^{16}$-dimensional spinorial wave function  ${\bm \Psi} = {{\Psi}}_{\sigma} (\beta^a ,  n^a_{\ i}, B_a)$
(where the spinorial index  $\sigma$ takes $2^{16}=65536$ values) that must satisfy the  $8 \times 2^{16}$ constraints
\begin{equation}
\label{susyconstraints}
\widehat {\cal S}^A \, \vert{\bm\Psi}\rangle = 0 , \quad \widehat { {\cal S}}^{A \dagger}  \, \vert{\bm\Psi}\rangle = 0 \,.
\end{equation}
Here, each of the $\widehat {\cal S}^A$'s and $\widehat { {\cal S}}^{A \dagger}$'s  is represented by
a $2^{16} \times 2^{16}$ matrix of first-order differential operators in the 
fourteen bosonic variables $\beta^a ,  n^a_{\ i}, B_a$. More precisely the structure of  $\widehat {\cal S}^A$ is
\bea \label{SA}
\widehat {\cal S}^A&=& \sum_a \widehat\pi_a \widehat \Phi^{aA} - \sum_{a<b}  e^{-(\beta^{ b}-\beta^{ a})}  \widehat{P}_{ab} 
(\gamma^{ab})^A_{\;\; B} (\widehat\Phi^{bB}-\widehat\Phi^{aB}) \nonumber\\
&-& i\frac{\sqrt{3}}{2} \sum_a e^{-\beta^a}\widehat P^a (\gamma^a)^A_{\;\; B}\widehat \Phi^{a B}+ \widehat {\cal S}^A_{(3)}\,,
\eea
where $\widehat\pi_a$, $\widehat{P}_{ab}$, $\widehat P^a $ are the first-order derivative operators defined
in Eqs. \eqref{qmomenta}, while the sixteen $\widehat \Phi^{aA}$ are $2^{16} \times 2^{16}$ ``gamma matrices"
satisfying the Clifford algebra \eqref{CRPhi}. The last term $\widehat {\cal S}^A_{(3)}$ in Eq. \eqref{SA} (which
is analogous to a matrix-valued mass term $\widehat M$  in a Dirac equation $ \gamma^{\mu} \widehat P_{\mu} \Psi + \widehat M \Psi=0$) is cubic in the  $\widehat \Phi^{aA}$'s and independent of bosonic degrees of freedom. It is defined by the ordering displayed in Eq. \eqref{qS3Phi}, with 
$\widehat {\cal S}^{ \dagger A}_{(3)}$ being
 correspondingly ordered. Note also that the momenta entering $\widehat {\cal S}^{ \dagger A}$ 
 contain ${\widehat\pi_a}^{\dagger}$, which is defined as usual
 as being ${\widehat\pi_a}^{\dagger} \equiv \widehat\pi_a$.

Similarly to the fact that the  first-order Dirac equation $ \gamma^{\mu} \widehat P_{\mu} \Psi + m \Psi=0$ entails the second-order Klein-Gordon equation $ \eta^{\mu \nu} \widehat P_{\mu} \widehat P_{\nu}\Psi + m^2 \Psi=0$, the
first-order (supersymmetry) constraints \eqref{SA} imply a quantum (Hamiltonian) constraint that is second-order
in the bosonic quantum momenta $ \widehat\pi_a$,  $ \widehat P_{ab}$,  $ \widehat P^a$. The ordering of this
quantum Hamiltonian constraint is fully determined by the above-defined ordering of the supersymmetry constraints.
When acting on the rescaled wave function
\be \label{Psiprime}
 {{\Psi}}_{\sigma}' (\beta^a , \varphi^b) = e^{+\varpi_a\beta^a} \, {{\Psi}}_{\sigma} (\beta^a , \varphi^b)\,,
\ee
the quantum Hamiltonian constraint reads
\be \label{Hprimebis}
 \widehat H_1^{\prime} {{\Psi}}' (\beta^a , \varphi^b)=0
\ee
where $\widehat H_1^{\prime}$ is a Klein-Gordon-like operator of the form
\bea \label{qH1bis}
\widehat H_1^{\prime}&=& G^{ab} \widehat\pi^{\prime}_a \widehat\pi^{\prime}_b + 2 \sum_{a<b} e^{-2(\beta^{ b}-\beta^{ a})} (\widehat {P}_{ab})^2  \nonumber\\ 
&& + \frac12 \sum_a e^{-2\,\beta^a}(\widehat P^a)^2  
 + \widehat H^{(2)} + \widehat \mu^2\,.
\eea
Here  $\widehat \pi^{\prime}_a \equiv \frac{\hbar}{i} \partial_{\beta^a}$ when  acting on $ {{\Psi}}_{\sigma}' (\beta^a , \varphi^b) $,  the bilinear coupling to the fermions $ \widehat H^{(2)}$ is given by
\be \label{qH2bis}
\widehat H^{(2)}= + 2 \sum_{a<b} e^{-(\beta^{ b}-\beta^{ a})}\widehat {P}_{ab} \widehat J_{ab}(\Psi) - \frac{1}{\sqrt{3}}\sum_a e^{-\beta^a}\widehat P^a \widehat J_a(\Psi)\,,
\ee
while the ``squared mass term"  $ \widehat \mu^2$ is quartic in the
fermions $\Phi$ and $\Phi^{\dagger}$, and independent of the bosonic degrees of freedom  $\beta^a ,  n^a_{\ i}, B_a$.
When one is far from all the walls (and on their positive sides), i.e. when all the linear forms $\beta^a$, and $\beta^{ b}-\beta^{ a}$ (with
$a<b$) are much larger than 1, one can neglect all the exponential terms,  so that the Hamiltonian
constraint reduces to a simple Klein-Gordon-like equation in the 4-dimensional $\beta$ space:
\be
\left(  G^{ab} \widehat\pi^{\prime}_a \widehat\pi^{\prime}_b + \widehat \mu^2 \right) {{\Psi}}'(\beta^a) =0\,.
\ee
However, the squared-mass term $ \widehat \mu^2 \equiv\widehat H_1^{\prime (4)}$ in the latter far-wall Klein-Gordon equation is not a c
number, but an operator in the quantum fermionic space, i.e. a  $2^{16} \times 2^{16}$ matrix acting
on the spinor index $\sigma$ of the wave function $ {{\Psi}}'_\sigma$.

%%%%%
\subsection{Kac-Moody structures in the quantum constraints}.

Having summarized  the quantum dynamics of our  five-dimensional supergravity cosmological model,
we can now highlight the hyperbolic Kac-Moody structures it contains. 

First, both the supersymmetry constraints,
and the Hamiltonian one, involve exponential terms of the form $e^{- \alpha_I(\beta)}$ (in $\widehat {\cal S}$
and  $\widehat {\cal S}^{\dagger}$) or  $e^{- 2\alpha_I(\beta)}$ (in $\widehat H_1^{\prime}$). Here,
the $\alpha_I(\beta)$'s are certain linear forms in the logarithmic scale factors $\beta^a$ parametrizing the
diagonal degrees of freedom of the spatial metric $h_{ij}$. There are ten such linear forms. Six of them, namely
\be
\alpha^{(ab)}(\beta) \equiv \beta^b-\beta^a \; ({\rm with} \; a<b)\,,
\ee
are called ``symmetry walls forms", and are linked to the off-diagonal degrees of freedom of the  spatial metric $h_{ij}(t)$,
while the remaining four ``electric wall forms", namely
\be
\alpha^{(a)}(\beta) \equiv \beta^a\,,
\ee
are linked to the time-dependent electric potential $A_i(t)$.
When endowing the 4-dimensional $\beta$ space with the Lorentzian-signature metric $G_{ab}$ defining the
kinetic terms of the $\beta^a(t)$'s, Eq.~\eqref{Gab}, the wall forms $\alpha^{(ab)}(\beta)$ and $\alpha^{(a)}(\beta)$
can be identified with real roots of the hyperbolic Kac-Moody algebra $G_2^{++}$. In addition, the four linear forms
$ \alpha^{(12)}(\beta),  \alpha^{(23)}(\beta),   \alpha^{(34)}(\beta),  \alpha^{(1)}(\beta)$ that can be
identified with the four {\it simple roots} of  $G_2^{++}$ are the ones that enter the four dominant potential walls
when considering the BKL-type chaos of  general
solutions of the (Einstein-Maxwell-like) bosonic dynamics of 5d supergravity near a cosmological singularity. 
[Indeed,
in the Weyl chamber defined by the positivity of $ \alpha^{(12)}(\beta),  \alpha^{(23)}(\beta),   \alpha^{(34)}(\beta),  \alpha^{(1)}(\beta)$, i.e. in the domain $0< \beta^1< \beta^2<\beta^3<\beta^4$, the other exponential
potentials are subdominant; e.g. as  $\alpha^{(13)}(\beta)=  \alpha^{(12)}(\beta)+  \alpha^{(23)}(\beta)$,
we have the subdominance property $e^{- \alpha^{(13)}(\beta)}=e^{- \alpha^{(12)}(\beta)}e^{- \alpha^{(23)}(\beta)}$.]

Besides the appearance of some of the roots of $G_2^{++}$, including the crucial simple roots (which suffice to
generate the full root lattice of $G_2^{++}$), the other Kac-Moody-related features exhibited
by  our quantum dynamics concern the fermionic sector. There are two such features. 

On the one hand,
the bilinear coupling to the fermions $ \widehat H^{(2)}$, Eq. \eqref{qH2bis}, associates to each one
of the wall roots $\alpha_I(\beta)=(\alpha^{(ab)}(\beta), \alpha^{(a)}(\beta))$, a coupling term of the generic form
\be
e^{- \alpha_I(\beta)} \widehat P_{\alpha_I} \widehat J_{\alpha_I}\,,
\ee
where $ \widehat P_{\alpha_I}$ is a quantum momentum associated with the bosonic variable $\alpha_I(\beta)$
(and contributing to the bosonic part of the Hamiltonian a term $\propto e^{- 2\alpha_I(\beta)} (\widehat P_{\alpha_I})^2$ ),  while $\widehat J_{\alpha_I}$ is a fermion bilinear. The important point here is that,
when normalizing\footnote{As discussed in Appendix B of \cite{Damour:2009zc} the appropriate Kac-Moody-related 
normalization of the momentum $P_{\alpha_I}$ depends on the squared-length 
$ \alpha_I^2=G^{ab} \alpha^I_a \alpha^I_b$ of the considered root. The normalization induced by the supergravity
dynamics happens to be appropriate  for a Kac-Moody interpretation.}
 the various fermion bilinears  $\widehat J_{\alpha_I}$ as in Eq. \eqref{defJq}, they do satisfy
the Serre-Berman relations Eq. \eqref{berman} as operators acting on the $2^{16}$-dimensional Clifford
representation space of the quantum fermions $\Phi$, $\Phi^{\dagger}$.
This follows from the fact that the second-quantization (for fermions, as is relevant here)
 has functorial properties in that it maps  classical generators $J_{\alpha}$ acting on some
 vectors $v$, members of some $n$-dimensional vector space ${\cal V}$,  onto quantum 
 operators  ${\widehat J}_{\alpha}$
 acting on the Fock  space built by piling up the successive antisymmetric powers of ${\cal V}$
 (up to the maximum power  ${\cal V}^{\wedge n}$ allowed by antisymmetry). [In our case, $n=16$ and 
 the space ${\cal V}$ is that of classical vector-spinors $v^{a A}$.] More precisely, given a linear endomorphism
$J_{\alpha}$  of  ${\cal V}$ (explicitly given, in some basis $e_i$ of  ${\cal V}$,
by a matrix $(J_{\alpha})^i_{\; j}$ acting on the vector index of $v = v^i e_i$, i.e. $(J_{\alpha} \cdot v)^i= 
(J_{\alpha})^i_{\; j} v^j$), the Fock space is $ {\mathbb C} \oplus {\cal V}  \oplus {\cal V}^{\wedge 2} \oplus \cdots {\cal V}^{\wedge n}$, and  the second-quantized ${\widehat J}_{\alpha}= \Phi^{\dagger}_i (J_{\alpha})^i_{\; j} \Phi^j$,
with $\{ \Phi^{\dagger}_i, \Phi^j \}= \delta_i^j$,
 decomposes as a direct sum of operators acting on each
(fermionic) level, from $N_F=0$, up to $N_F=n$. More precisely:  at level $ N_F=0$ 
(Fock vacuum, $ \vert 0\rangle_-$), ${\widehat J}_{\alpha}$
acts like 0; at level $N_F=1$, ${\widehat J}_{\alpha}$ acts on ${\cal V}$  like ${ J}_{\alpha}$;
at level $N_F=2$, ${\widehat J}_{\alpha}$ acts on ${\cal V}^{\wedge 2}$  like 
\be \label{Jlevel2}
{\widehat J}_{\alpha} |^{N_F=2} = (J_{\alpha} \otimes \mathbbm{ 1}) \oplus ( \mathbbm{ 1} \otimes J_{\alpha} )\,.
\ee
Explicitly, the meaning of the latter equation is that ${\widehat J}_{\alpha} |^{N_F=2}$ acts on a 
(factorized\footnote{A generic element of ${\cal V}^{\wedge 2}$ is a linear combination of such factorized elements.})
element $u \wedge v \in {\cal V}^{\wedge 2}$ as $ (J_{\alpha} \cdot u) \wedge v + u \wedge ( J_{\alpha} \cdot v)$. At the
fermionic level $N_F$, ${\widehat J}_{\alpha}$ decomposes as a sum of $N_F$ terms of the same type as
indicated in Eq. \eqref{Jlevel2}, e.g. 
\be \label{Jlevel3}
{\widehat J}_{\alpha} |^{N_F=3} = (J_{\alpha} \otimes \mathbbm{ 1} \otimes \mathbbm{ 1}) \oplus ( \mathbbm{ 1} \otimes J_{\alpha} \otimes \mathbbm{ 1} ) \oplus  ( \mathbbm{ 1} \otimes \mathbbm{ 1} \otimes J_{\alpha}). 
\ee
This nice functorial nature of the map transforming a classical operator $J_{\alpha}$ into a corresponding
second-quantized one $\widehat J_{\alpha}$ allows one to transport many  properties
satisfied by $J_{\alpha}$ into corresponding properties of $\widehat J_{\alpha}$. 

For instance, classical commutators $ [J_{\alpha_1}, J_{\alpha_2}]$ are mapped onto
 their corresponding quantum ones, namely
 \be \label{functor}
   [\widehat J_{\alpha_1}, \widehat J_{\alpha_2}]= \widehat{[J_{\alpha_1}, J_{\alpha_2}]}\,.
  \ee
 This functorial property ensures, in particular, that, if  we have, say,  $ [J_{\alpha_1}, J_{\alpha_2}]= c J_{\alpha_3}$,
 the corresponding quantum commutators satisfy $ [\widehat J_{\alpha_1}, \widehat J_{\alpha_2}]= c \widehat J_{\alpha_3}$. This guarantees, in particular, that Serre-Berman relations Eq. \eqref{berman} are preserved 
 by the quantization. An important consequence is that
 the root operators $\widehat J_{\alpha_I}$ entering the quantized
 Hamiltonian $\widehat H^{(2)}$ generate a $2^{16}$-dimensional representation of $K(G_2^{++})$,
 the maximally compact subalgebra of $G_2^{++}$ fixed by the Chevalley involution.
 We will indicate below another important consequence of these functorial properties concerning
 the reflection operators of quantum fermions in the short-wavelength limit of the cosmological dynamics.

In addition,  we have also explicitly proven that the term quartic in fermions in the quantum
Hamiltonian constraint, namely $\widehat \mu^2$ in Eq. \eqref{qH1bis}, {\it commutes} with all the root operators
$\widehat J_{\alpha_I}$:
\be
[\widehat J_{\alpha_I}, \widehat \mu^2]=0\; ; \; {\rm for} \; I= (ab), (a)\,.
\ee
Quite remarkably, the latter commutation property is rooted in a hidden simple structure of the
quartic-in-fermion term. Indeed, we found that $\widehat \mu^2$ can be expressed in
terms of two simple fermion-bilinears $\widehat N_F$ and $\widehat C_F$, which separately commute with the root operators
$\widehat J_{\alpha_I}$. Namely, 
\bea \label{mu2}
\widehat \mu^2 &=& \frac{14}{3} -\frac12 (\widehat N_F-8)^2 -\frac14 (\widehat C_F^{\dagger} \widehat C_F+ \widehat C_F \widehat C_F^{\dagger}) \nonumber\\
&=& -\frac{106}{3} + 9 \widehat N_F - \frac12 \widehat N_F^2 -\frac12 \widehat C_F^{\dagger} \widehat C_F\,,
\eea
with
\be \label{qNF}
\widehat N_F \equiv G_{ab} \widehat \Phi^{a  \dagger}  \widehat \Phi^{b } \equiv G_{ab} \widehat \Phi^{a A \dagger} \delta_{AB} \widehat \Phi^{b B}\,,
\ee
and
\be \label{qCF}
\widehat C_F \equiv G_{ab} \widehat \Phi^{a A} C_{AB} \widehat \Phi^{b B}\,.
\ee
Eq.~\eqref{qNF} defines the quantum fermion number, with eigenvalues $N_F=0,1,\cdots,16$.
In Eq.~\eqref{qCF}  the $4 \times 4$ matrix $C_{AB}$ is the ``charge conjugation" matrix of the (spatial) 
$\gamma_i$ matrices, defined so that it is hermitian, $C^{\dagger}= C$,
and satisfies $C \gamma_i C^{-1}= - \gamma_i^T$. [$C_{AB}$ is an antisymmetric
matrix in all representations of the $\gamma$ matrices.] We then have
\be
\widehat C_F^{\dagger} \equiv G_{ab} \widehat \Phi^{b B \dagger} C_{AB} \widehat \Phi^{a A\dagger}
= -  G_{ab} \widehat \Phi^{a A \dagger} C_{AB} \widehat \Phi^{b B \dagger}\,.
\ee
As already said, both $\widehat N_F$ and  $\widehat C_F$ (and therefore also $\widehat C_F^{\dagger}$)
commute with all the $\widehat J_{ab}$'s and  $\widehat J_{a}$'s. Note that while $\widehat N_F $
is a sesquilinear form  $\widehat N_F  \sim \Phi^{\dagger} \Phi$ that is hermitian,  $\widehat C_F$ is
a symplectic bilinear form in the $\Phi$'s (which would vanish if the $\Phi$'s would commute rather than anticommute).
It is also to be noted that
\be
\widehat N_F-8 = \frac12G_{ab}  \left(  \widehat \Phi^{a  \dagger}  \widehat \Phi^{b }-  \widehat \Phi^{a }  \widehat \Phi^{b  \dagger}  \right)
\ee
is  odd under the up-down fermion symmetry where one swaps $ \Phi \leftrightarrow \Phi^{\dagger}$.
The first line in Eq. \eqref{mu2} then shows that $\widehat \mu^2$ is also invariant under the swapping $ \Phi \leftrightarrow \Phi^{\dagger}$. [The up-down fermion symmetry was used above as part of our definition
of the ordering of the supersymmetry constraints.]

From the mathematical point of view, as already mentioned above,
any four operators $J_1$, $J_{12}$, $J_{23}$, $J_{34}$
(acting as endomorphisms of some vector space)
satisfying the Serre-Berman relations Eq. \eqref{berman} define a representation of the 
(formally) maximally compact subalgebra $K(G_2^{++})$ of $G_2^{++}$. 
We can therefore summarize the results of the present section by saying that
the fermions of our quantized supersymmetric cosmological model live in a $2^{16}$-dimensional representation
of $K(G_2^{++})$, and that all the building blocks entering the dynamics of the fermions, i.e. the
various terms defining $\widehat H^{(2)} \sim \Phi^{\dagger} \Phi$ and 
 $\widehat H^{(4)} \sim \Phi^{\dagger} \Phi  \Phi^{\dagger} \Phi$ have a direct meaning in terms of the
 simple-root generators $\widehat J_1$, $\widehat J_{12}$, $\widehat J_{23}$, $\widehat J_{34}$
 of $K(G_2^{++})$.
 
 %%%%
 \section{Solutions of the quantum constraints}

In this final section, we briefly discuss some aspects of the solutions of our quantized cosmological
model, i.e. the solutions of the supersymmetry constraints \eqref{susyconstraints}. We recall that
the latter supersymmetry constraints entail the Hamiltonian constraint, say \eqref{Hprimebis}.

Let us first focus on the structure of the solutions far from all the walls, i.e. in a domain of the $\beta^a$'s
where we can neglect all the exponential terms $ e^{-(\beta^{ b}-\beta^{ a})}$ and  $e^{-\beta^{ a}}$
in the $\widehat S_A$'s, and their squares in $\widehat H$.  In this limit the supersymmetry constraints
reduce to
\be \label{SAbis}
\left(  \widehat \Phi^{aA} \frac{\hbar}{i} \, \partial_{\beta^a} + \widehat {\cal S}^A_{(3)} \right) \vert{\bm\Psi (\beta)}\rangle = 0\,,
\ee
\be \label{SAdaggerbis}
\left(  \widehat \Phi^{aA \dagger} \frac{\hbar}{i} \, \partial_{\beta^a} + \widehat {\cal S}^{A \dagger}_{(3)} \right) \vert{\bm\Psi (\beta)}\rangle = 0\,,
\ee
while the Hamiltonian constraint reads
\be
\left[ G^{ab}\left( \frac{\hbar}{i} \, \partial_{\beta^a} - i \varpi_a \right) \left( \frac{\hbar}{i} \, \partial_{\beta^b} - i \varpi_b \right) + \widehat \mu^2\right] \vert{\bm\Psi (\beta)}\rangle = 0\,,
\ee
where $\varpi_a \equiv G_{ab}\varpi^a =- \frac 14\{9,8,7,6\}$. In these equations we have formally
considered that the operator $\widehat \pi_a$ was hermitian, and we have used the original, non-rescaled
wavefunction $\Psi(\beta)$ (rather than the rescaled wavefunction $\Psi'(\beta)$, Eq. \eqref{Psiprime},
 used in Eq. \eqref{Hprimebis}).

 %%%%%%%%
\subsection{Spectrum of $\widehat \mu^2$ }

 To solve the Hamiltonian constraint we can look for solution states $\vert{\bm\Psi}\rangle$ that are
 eigenstates of the $ \widehat \mu^2$ operator. It is therefore interesting to first discuss the eigenvalues
 and eigenstates (in fermionic space) of $ \widehat \mu^2$. The explicit expression \eqref{mu2} of $ \widehat \mu^2$
 show that $ \widehat \mu^2$ commutes with $\widehat N_F$. The latter operator defines the fermion number
with respect to the Fock vacuum of the $\Phi$'s, i.e. the empty state $ \vert 0\rangle_-$ such that
\be
 \widehat \Phi^{aA}  \vert 0\rangle_-=0.
\ee

Starting from this empty state, the $ N_F=1$ states are obtained by acting on $ \vert 0\rangle_-$
with any of the sixteen anticommuting fermionic creation operator $ \widehat \Phi^{aA \dagger}$, etc.
The number of states at level $N_F$ is then equal to ${16 \choose  N_F}= {16 \choose 16- N_F}$, i.e. 16 for 
$N_F=1$  (and $N_F=15$),
120 for $N_F=2,14$, etc., with a maximum value $ {16 \choose  8} =12870$ for $N_F=8$.
The filled state,  say $ \vert 0\rangle_+$,  at level $N_F=16$ is unique and such that
\be
 \widehat \Phi^{aA \dagger}  \vert 0\rangle_+=0.
\ee
  The explicit expression \eqref{mu2} of $ \widehat \mu^2$ allows one to prove that
 $ \widehat \mu^2$ also commutes with the operators $\widehat C_F$ and $\widehat C_F^{\dagger}$:
 \be
 [\widehat \mu^2, \widehat C_F] =0\; ; \;  [\widehat \mu^2, \widehat C_F^{\dagger}] =0\,.
 \ee
This is seen by using the easily checked commutation relations
\eqal{
&\big[\widehat N_F,\,\widehat { C_F}\big]=-2\,\widehat { C_F} \,,\\
&\big[\widehat N_F,\,{\widehat {  C_F}}{}^\dagger\big]=+ 2 \,\widehat{ C_F}{}^\dagger \,,\\
&\big[ \widehat { C_F}{}^\dagger\,,\widehat { C_F}\big]=+4\,\widehat N_F-32 \,.
}
Noting that $\widehat C_F^{\dagger}$ increases the value of $N_F$ by 2, while  $\widehat C_F$
decreases $N_F$ by 2, and that they both commute with $\widehat \mu^2$, we can use 
 $\widehat C_F^{\dagger}$ and  $\widehat C_F$ as ladder operators to map some sub-eigenspaces
of  $\widehat \mu^2$ at fermion level $N_F$ onto corresponding eigenspaces of $\widehat \mu^2$
at fermion levels $n_F \pm 2$, with the same value of $\mu^2$.
This yields the following spectrum of $\widehat \mu^2$ when $N_F$ varies between 0 and 8 (with symmetric
results when $N_F'= 16 - N_F$)
\begin{widetext}
\eqals{
N_F=0, 16\qquad&\mu^2=- \frac {106}{3}\big\vert_1\\
N_F=1, 15\qquad&\mu^2=-\frac {161}6\big\vert_{16}\\
N_F=2, 14\qquad&\mu^2= - \frac {106}{3}\big\vert_{1} \quad&,&\quad -\frac{58}3 \big\vert_{119}&\ &\ &\ \\
N_F=3, 13\qquad&\mu^2= -\frac {161}6\big\vert_{16} \quad&,&\quad -\frac{77}6 \big\vert_{544}&\ &\ &\ \\
N_F=4, 12\qquad&\mu^2=- \frac {106}{3}\big\vert_{1}  &,&\quad -\frac{58}3\big\vert_{119}\quad &,&\quad -\frac{22}3\big\vert_{1700}&\ &\ \\
N_F=5, 11\qquad&\mu^2= -\frac {161}6\big\vert_{16}\quad &,&\quad -\frac{77}6\big\vert_{544}\quad&,&\quad-\frac{17}6\big\vert_{3808}&&\\
N_F=6, 10\qquad&\mu^2=-\frac{106}3\big\vert_{1}\quad &,&\quad- \frac{58}3\big\vert_{119}\quad &,&\quad-\frac{22}3 \big\vert_{1700}\quad &,&\quad +\frac{2}3\big\vert_{6188}&\\
N_F=7, 9\qquad &\mu^2=  -\frac {161}6\big\vert_{16}\quad &,&\quad -\frac{77}6\big\vert_{544}\quad &,&\quad-\frac{17}6 \big\vert_{3808}\quad &,&\quad+\frac{19}6\big\vert_{7072}&\\
N_F=8\qquad &\mu^2=-\frac{106}3\big\vert_{1}\quad &,&\quad -\frac{58}3\big\vert_{119}\quad &,&\quad-\frac{22}3 \big\vert_{1700}\quad &,&\quad +\frac{2}3\big\vert_{6188}\quad&,&\quad+\frac{14}3\big\vert_{4862}
}
\end{widetext}
Here the numbers indicated after the eigenvalues of $\widehat \mu^2$ denote the dimensions of the
corresponding eigenspaces. For instance, the one-dimensional eigenspace $\mu^2= - \frac {106}{3}$
at level $N_F=2$ is obtained by acting on the unique $N_F=0$ state by $\widehat { C_F}{}^\dagger$.
In other words, if we define the function 
\be
f(n) \equiv  -\frac{106}{3} + 9 n - \frac12 n^2 \,,
\ee
the possible eigenvalues of $\widehat \mu^2$ at a given level $N_F$ are of the form
$f(n)$, with degeneracy ${16 \choose n}- {16 \choose n-2}$, where the integer $n$ runs over
the values $N_F, N_F-2, N_F-4, \cdots$.

%%%%%%%%
\subsection{Far-wall solutions of the quantum constraints at low (and high) fermion levels}

The above-determined spectrum of  $\widehat \mu^2$ yields a necessary constraint on possible solution wavefunctions,
but is far from sufficient to determine whether such solutions exist at some given fermion level $N_F$.
[The reasoning given below Eq. (11.21) of Ref. \cite{Damour:2014cba} shows that one can look for solutions having
a given $N_F$ level.] We must tackle the supersymmetry constraints, Eqs.  \eqref{SAdaggerbis},   \eqref{SAdaggerbis}.
We succeeded in doing so for the levels $N_F=0,1,2,3$ and their up-down symmetric partners  $N_F=16,15,14,13$. 

The main result at the levels $N_F=0,1,2,3$ (and $N_F=16,15,13$) is that there exist solutions 
of the type 
\be
\vert{\bm\Psi (\beta)}\rangle = \exp (i \pi_a \beta^a) \vert{\bm\Psi(0) }\rangle\,,
\ee
 only for certain specific, discrete values of the momenta $\pi_a $.

At the level $N_F=0$, $\vert{\bm\Psi(0) }\rangle$ must be proportional to $ \vert 0\rangle_-$,
while  $\pi_a $ must take the specific value  $\pi_a^{N_F=0}=\frac i4 \{19,16,13,10\}$.
Note that the corresponding value $\pi_a' $ parametrizing the rescaled wavefunction 
$\vert{\bm\Psi' (\beta)}\rangle $, namely
\be
\pi_a'= \pi_a- i \varpi_a\,,
\ee
is also purely imaginary and is fixed to the specific value
\be
{\pi'}_a^{N_F=0}= i  \{7,6,5,4\}\,.
\ee
It is easily checked that $G^{ab} {\pi'}_a^{N_F=0}{ \pi'}_b^{N_F=0}$ is equal to $- \mu^2_{N_F=0}= + \frac{106}{3}$,
as it should be.

At the level $N_F=1$, we found that there does not exist any solution of the supersymmetry constraints.

At the level $N_F=2$, there exist only five possible, discrete values of the momenta $\pi_a$, all of them
being purely imaginary. The corresponding linear space of solutions is 6-dimensional, because one value
of $\pi_a$ (namely $\pi_a^{(1)}=\frac i4 \{11,8,9,6\}$) admits a 2-dimensional space of solutions for
the spinor factor
$\vert{\bm\Psi(0) }\rangle$. The other possible values of $\pi_a$ at $N_F=2$ are:
$\pi_a^{(2)}= \frac i4 \{19,16,13,10\}= \pi_a^{N_F=0}$ (with spinor part  $\widehat{\mathcal C_F}{}^\dagger \vert 0\rangle$), $\pi_a^{(3)}=\frac i4 \{7,12,9,6\}$, $\pi_a^{(4)}=\frac i4 \{11,8,5,10\}$ and $\pi_a^{(5)}=\frac i4 \{23,12,9,6\}$. 
The values of $\mu^2$ corresponding to the five possible momenta at level $N_F=2$
are $(\mu^2)^{(1)}=(\mu^2)^{(3)}=(\mu^2)^{(4)}=(\mu^2)^{(5)}= - \frac{58}{3} $
and $(\mu^2)^{(2)}= - \frac{106}{3}$.

At the level $N_F=3$, there exists only one possible, discrete value of $\pi_a$, namely 
 $\pi_a^{N_F=3}=\frac i4 \{5,8,3,6\}$ (with $ \mu^2_{N_F=3}= - \frac{77}{6}$), 
 with a corresponding 4-dimensional eigenspace for the spinor part  $\vert{\bm\Psi(0) }\rangle$.

 There exist corresponding mirror solutions at $N_F=16,14,13$ 
 with correspondingly equal  values of $\pi_a$.
 More generally the up-down symmetry in fermion space guarantees that one can map any solution 
 at any level $N_F$ into a corresponding solution at level $16-N_F$. Indeed, under the transformation
 where\footnote{Here, we omit for simplicity the hats on the various quantum operators.}
  $\Phi \mapsto {\widetilde \Phi} \equiv \Phi^{\dagger }$ (and therefore
  $\Phi^{\dagger } \mapsto {\widetilde \Phi}^{\dagger } \equiv \Phi$)
  our ordering, Eq. \eqref{qS3Phi}, shows that ${\cal S} \mapsto  \widetilde {\cal S}$,
 where  $\widetilde {\cal S}\sim \pi \widetilde \Phi + c  \widetilde \Phi \widetilde \Phi^{\dagger} \widetilde \Phi $
 is simply equal ${\cal S}^{\dagger}$. Then, using   Eq. \eqref{updown1} (and $(\widehat \pi)^{\dagger} \equiv \widehat \pi$), one finds that  ${\cal S}^{\dagger} \mapsto  \widetilde {\cal S}^{\dagger}$, where
 $\widetilde {\cal S}^{\dagger}$ is  simply equal to  ${\cal S}$. Thereby any solution $\vert{\bm\Psi (\beta)}\rangle$ at some level $N_F=n$ constructed by acting on the  empty state $ \vert 0\rangle_-$ with $n$
 creation operators  $\Phi^{\dagger \cal A}= \Phi^{\dagger a A} $, say
 \be \label{Psin}
 \vert{\bm\Psi (\beta)}\rangle = X_{{\cal A}_1 {\cal A}_2 \cdots {\cal A}_n}(\beta) \Phi^{\dagger {\cal A}_1} 
  \Phi^{\dagger {\cal A}_1} \cdots  \Phi^{\dagger {\cal A}_n} \vert 0\rangle_-\,,
 \ee
 with coefficients $ X_{{\cal A}_1 {\cal A}_2 \cdots {\cal A}_n}(\beta)= X_{[{\cal A}_1 {\cal A}_2 \cdots {\cal A}_n]}(\beta)$,
 can be automatically mapped into a  corresponding mirror solution at level $N_F=16-n$ obtained
 by acting on the filled state $ \vert 0\rangle_+$ (which is annihilated by the ${\widetilde \Phi}^{ \cal A}$'s)
 with the operators ${\widetilde \Phi}^{\dagger \cal A} \equiv \Phi^{ \cal A}$,
 namely
  \be
 \vert{\bm \widetilde \Psi (\beta)}\rangle = X_{{\cal A}_1 {\cal A}_2 \cdots {\cal A}_n }(\beta) 
 \Phi^{ {\cal A}_1} \Phi^{ {\cal A}_2} \cdots  \Phi^{ {\cal A}_n} \vert 0\rangle_+\,.
 \ee
 Note that this mirror solution at level $16-n$ involves the same coefficients $ X_{{\cal A}_1 {\cal A}_2 \cdots {\cal A}_n }(\beta)$. In particular, when considering plane-wave solutions, 
 $ X_{{\cal A}_1 {\cal A}_2 \cdots {\cal A}_n }(\beta)= e^{i \pi_a \beta^a}X_{{\cal A}_1 {\cal A}_2 \cdots {\cal A}_n }(0) $, this up-down symmetry maps a momentum $\pi_a$ at level $n$ into the same
 momentum $\pi_a$ at level $16-n$.

 In addition to this up-down symmetry of the space of solutions of the constraints, there is an
 additional $\mathbb Z_2$ symmetry mapping any solution at level $N_F$ into a corresponding
 solution at the same level. This second symmetry is rooted in the reality structure of the
 supersymmetry constraints, namely in the fact that the numerical coefficients 
 $  c^A_{  {\cal A} {\cal B}  {\cal C}}$ entering
Eq. \eqref{qS3Phi} are  purely imaginary (in a suitable quasi-Majorana representation).
Indeed, when decomposing the supersymmetry constraints  
$ {\cal S}\vert{\bm\Psi (\beta)}\rangle =0$, $ {\cal S}^{\dagger}\vert{\bm\Psi (\beta)}\rangle =0$,
with a state of the form Eq. \eqref{Psin}, on the Fock states 
$\Phi^{\dagger {\cal A}_1} 
  \Phi^{\dagger {\cal A}_1} \cdots  \Phi^{\dagger {\cal A}_k} \vert 0\rangle_-$
  at levels $k=n-1$ and $k=n+1$, one gets a system of first-order differential equations for the coefficients 
  $X_{{\cal A}_1 {\cal A}_2 \cdots {\cal A}_n }(\beta)$ of the symbolic form (using $\hbar=1$)
  \be \label{explicitsusy}
  \frac{1}{i} \, \partial_{\beta} X(\beta) + G \,\delta\,  c \, X(\beta)=0\,.
  \ee
Here, the numerical coefficients $\sim  G \,\delta\,  c$ coming from the cubic-in-fermions 
contributions involve 
the  coefficients $c =  c^A_{  {\cal A} {\cal B}  {\cal C}}$ entering
Eq. \eqref{qS3Phi}, multiplied  by the real coefficients  $ G^{ab} \, \delta^{AB}$ coming from the use of the
anticommutation relations  Eqs. \eqref{CRPhi}. The explicit form of the supersymmetry constraints, 
Eq. \eqref{explicitsusy}, are given in Appendix \ref{susyex}.
Using the pure-imaginary nature of the  $c^A_{  {\cal A} {\cal B}  {\cal C}}$'s, we see that (after multiplying them by $i$) the supersymmetry constraint equations 
yield a system of  {\it real} partial differential equations for the wavefunction
 $X_{{\cal A}_1 {\cal A}_2 \cdots {\cal A}_n }(\beta)$. Therefore, to any given  (generally) complex
 solution  $X_{{\cal A}_1 {\cal A}_2 \cdots {\cal A}_n }(\beta)$ at level $n$, one can associate 
 a solution having the complex-conjugated wavefunction  
 $X^*_{{\cal A}_1 {\cal A}_2 \cdots {\cal A}_n }(\beta)$. For instance, under this map a plane-wave
 solution of momentum $\pi_a$ at level $n$ is transformed into a corresponding
 solution at the same level with momentum  $ -\pi_a^*$. For generic solutions at the intermediate
 levels $4 \leq n \leq 12$ such an involutory map acts non trivially on the space of solutions.
 On the other hand, it acts trivially on the solutions discussed above at levels $n=0,2,3$ and $n=13,14,16$,
 which are purely real (up to an arbitrary overall complex factor).

 %%%%%%%%
\subsection{Short-wavelength continuous far-wall solutions of the quantum constraints for $4 \leq N_F \leq 12$}
 
 It was found in the study of the quantum cosmological dynamics of 
$D=4$, $N=1$ supergravity \cite{Damour:2014cba,Damour:2017cpi}, that continuous solutions of the
supersymmetry constraints (with real $\pi'_a$'s taking
all possible values on its allowed  mass-shell $G^{ab} \pi'_a \pi'_b=- \mu^2$)
exist only in the middle of fermionic space, namely $N_F=2,3,4$. These solutions
were also shown to be continuously connected to their short-wavelength analogs,
obtained by taking the limit $\pi'_a \gg 1$. In the latter limit, one can neglect the cubic term 
 $ \widehat {\cal S}^A_{(3)}$ in the supersymmetry constraint, and the corresponding finite
 value of the quartic term $\mu^2=O(\hbar^2)$. We shall here assume that such a general feature
 holds also in our present $D=5$, $N=2$ supergravity case.
 
 Under this (plausible) assumption, we can complete our explicit study of the discrete solutions
 existing at low (and high) values of $N_F$ by delineating the general structure of the
continuous-$\pi'$ solutions existing for the remaining values, namely $4 \leq N_F \leq 12$.
[There might also exist additional discrete solutions; e.g., related by the ladder operators
$\widehat C_F$, $\widehat C_F^{\dagger}$, to the discrete solutions discussed above.]

When considering,  short-wavelength states, 
$\vert{\bm\Psi (\beta)}\rangle = \exp (i \pi_a \beta^a) \vert{\bm\Psi(0) }\rangle$,
with $\pi_a \gg1$, or equivalently, for the rescaled wavefunction
$\vert{\bm\Psi' (\beta)}\rangle = \exp (i \pi'_a \beta^a) \vert{\bm\Psi(0) }\rangle$
with $\pi_a'= \pi_a- i \varpi_a$, the supersymmetry constraints yield
\be
 \widehat \Phi^{aA} \pi'_a  \vert{\bm\Psi (0)}\rangle = 0,
\ee
\be
 \widehat \Phi^{aA \dagger} \pi'_a  \vert{\bm\Psi (0)}\rangle = 0,
\ee
which imply the (Hamiltonian-constraint) consequence
\be
G^{ab}  \pi'_a  \pi'_b=0\,.
\ee
Let us associate to any real (co)vector $v_a$ in (the dual of the) $\beta$ space the fermionic operators
(putting the spinor index $A$ down for convenience)
\be
 \widehat \Phi^v_A \equiv v_a \widehat \Phi^{a}_{A} \; ; \; \widehat \Phi^{v \dagger}_A \equiv v_a \widehat \Phi^{a \dagger}_{A}.
\ee
Given two covectors $u$ and $v$, the so-defined fermionic operators satisfy the (Clifford) relations
\be \label{cliff}
\{ \widehat \Phi^u_A,  \widehat \Phi^{v \dagger}_B \}= u \cdot v \,\delta_{AB}\,,\, \{ \widehat \Phi^u_A,  \widehat \Phi^{v}_B \}=0\,, \{ \widehat \Phi^{u \dagger}_A,  \widehat \Phi^{v \dagger}_B \}=0,
\ee
where $ u \cdot v \equiv G^{ab} u_a v_b$.

Given some $\pi_a'$ on the (Hamiltonian-constraint) light cone $ \pi'^2= G^{ab}  \pi'_a  \pi'_b=0$,
we can complete  $\pi_a'$ into a null frame  $\pi_a'$, $n_a$, $t^1_a$,  $t^2_a$ in the (dual) 
4-dimensional Lorentzian $\beta$ space. Here,  $\pi_a'$, $n_a$ are both null,
$0= \pi'^2=n^2$ (with the relative normalization $\pi' \cdot n = 1$),
while the two complementary vectors $t^1_a$,  $t^2_a$ are {\it transverse} to the null direction
$\pi'_a$, i.e. satisfy $0=\pi' \cdot t^1 = \pi' \cdot t^2 $. One can also require that $t^1$ and $t^2$ are
orthogonal to $n$, and between themselves, and (being necessarily spacelike) are normalized to unity.
From the basic relations \eqref{cliff}, and the fact that the supersymmetry constraints
read $\widehat \Phi^{\pi'}_{ A}   \vert{\bm\Psi (0)}\rangle = 0$,
 $\widehat \Phi^{\pi'\dagger}_{ A}    \vert{\bm\Psi (0)}\rangle = 0$,
one easily sees that the lowest value of $N_F$ where there can exist a short-wavelength
solution is $N_F =4$, and that, for this value, there is, for any given (null) $\pi'$
 a one-dimensional space of solutions of the type
 \be \label{N4}
 C  \exp (i \pi'_a \beta^a) \widehat \Phi^{\pi'\dagger}_{ 1} \widehat \Phi^{\pi'\dagger}_{ 2} \widehat \Phi^{\pi'\dagger}_{ 3} \widehat \Phi^{\pi'\dagger}_{ 4}\vert{0 }\rangle_-\,.
 \ee
 Then, at the $N_F=5$ level, there will be (for any given null $\pi'$) a eight-dimensional
 space of solutions generated by acting on the state in Eq. \eqref{N4} with any of the eight
 independent raising operators $ \widehat \Phi^{t^1\dagger}_{ A}$, and  $ \widehat \Phi^{t^2\dagger}_{ B}$,
 involving the two transverse vectors $t^1$ and $t^2$ constructed above. At the  $N_F=6$ level,
 there will be a $\frac{8 \times 7}{2}=28$-dimensional space of solutions obtained by acting 
 on the state in Eq. \eqref{N4} with a product of two raising operators of the form 
  $ \widehat \Phi^{t^1\dagger}_{ A}$, or  $ \widehat \Phi^{t^2\dagger}_{ B}$. One can
  continue generating such solutions up to the maximum value $N_F=12$, corresponding to
  acting on the state in Eq. \eqref{N4} with the eight different operators 
  $ \widehat \Phi^{t^1\dagger}_{ A}$, or  $ \widehat \Phi^{t^2\dagger}_{ B}$.

  %%%%%%%%
\subsection{Reflection of short-wavelength  solutions on potential walls}

Let us finally briefly discuss another consequence of our assumption that there exist 
solutions of the quantum supersymmetry constraints that are continuously
connected to the approximate solutions which one  obtains by working
in the Wentzel-Kramers-Brillouin (WKB), short-wavelength  approximation.
This approximation being the quasi-classical approximation ($\hbar \to 0$),
we further expect that such solutions will also correspond to the approximation
where the spin degrees of freedom are described by anticommuting Grassmann
variables ($ \{ \Phi^{\dagger}, \Phi\}=0$) rather than (as we did above) by quantum operators
satisfying a Clifford-algebra relation $ \{ \Phi^{\dagger}, \Phi\}=O(\hbar)$.

In the Grassmann-fermion approximation, it was generally shown (even in the non-simply-laced
case of relevance here) in Ref. \cite{Damour:2009zc} that  the law of evolution
of a fermion field $\Phi^i$ (where we use here, for generality, a generic index $i$ to label the 
representation space in which lives the considered fermion field) under Hamiltonians containing, in addition
to the usual Toda-like bosonic dynamics,
\be
H^{(0)} = \frac12 G^{ab} \pi_a \pi_b +  \sum_{I} e^{- 2\alpha_I(\beta)}  P_{\alpha_I}^2 \,,
\ee
 fermion couplings of the related Toda-type, namely
\be
H^{(2)} \approx \sum_{I} e^{- \alpha_I(\beta)}  P_{\alpha_I} \widehat J_{\alpha_I}\,,
\ee
where
\be
\widehat J_{\alpha_I}= \Phi^{\dagger}_i  (J_{\alpha_I})^i_{\; j }\Phi^j \,,
\ee
could be approximately integrated, and led to a ``Fermionic Billiard" picture.
More precisely, the latter Fermionic-Billiard picture is based on the fact that
the approximate  integration of the law of evolution of the fermion field
near each separate wall\footnote{The billiard approximation consists in treating both
the bosonic and the fermionic dynamics as a free far-wall evolution interrupted by
time-localized interactions with well-separated potential walls.}, namely
\be
\partial_t \Phi^i \approx i \, e^{- \alpha_I(\beta)} P_{\alpha_I}  (J_{\alpha_I})^i_{\; j }\Phi^j \,,
\ee
leads to a transformation linking the incident value
of the Grassmann-valued $\Phi^i$ to its reflected value given by a classical, fermionic reflection operator of the form
\be\label{RalphaG}
\mathcal{ R}_{\alpha_I}^{\rm classical} = e^{ i  \frac{\pi}{2} \varepsilon_{\alpha_I}  J_{\alpha_I}} \,,
\ee
where $\varepsilon_{\alpha_I} = \pm$ denotes the sign of the momentum $P_{\alpha_I}$.
In Eq. \eqref{RalphaG},  $ J_{\alpha_I}$ denotes the matrix  $(J_{\alpha_I})^i_{\; j }$
acting on the representation space defined by a classical homogeneous gravitino, and the
resulting classical reflection operator $\mathcal{ R}_{\alpha_I}^{\rm classical}$ (obtained by
exponentiating $(J_{\alpha_I})^i_{\; j }$ is also a matrix (or endomorphism) in the
representation space of the classical (i.e. Grassmannian) fermion field $\Phi^i$.

When working, as we do here, with second-quantized fermions, i.e. when replacing
the Grassmann fermion field $\Phi^i$ by a linear operator $\widehat \Phi^i$ acting
in a fermionic Fock space, we can  use the functorial character of the 
Fock-type second quantization (illustrated in our case by the definition, Eq.~\eqref{defJq},
of the second-quantized $\widehat J_{\alpha_I}$, and the fact that they have the same
algebraic properties as their first-quantized analogs, $ (J_{\alpha_I})^i_{\; j }$)
to map the classical reflection matrix $\mathcal{ R}_{\alpha_I}^{\rm classical}$
onto a corresponding reflection operator acting in the representation space of
the quantized fermion. 

In other words, under our assumption that the quasi-classical limit of our
quantum supersymmetric cosmological model does continuously connect 
quantum states to quasi-classical states, we conclude that, in the
 short-wavelength limit, the spinor factor,  $\vert{\bm\Psi (0)}\rangle$
 (stripped of the plane-wave factor $\exp (i \pi'_a \beta^a)$),
 of the quantum plane-wave solution states discussed
 in the previous subsection,
 \be
 \vert{\bm\Psi' (\beta)}\rangle = \exp (i \pi'_a \beta^a) \vert{\bm\Psi(0) }\rangle
 \ee
 (see, e.g., Eq. \eqref{N4} in the $N_F=4$ subspace),
 considered as states in the $2^{16}$-dimensional representation space of the quantized
 gravitino, will be transformed, upon reflection on each (symmetry or electric)
 potential wall under the quantum reflection operator
\be\label{Ralphaq}
\mathcal{ R}_{\alpha_I}^{\rm quantum} = e^{ i  \frac{\pi}{2} \varepsilon_{\alpha_I} \widehat J_{\alpha_I}} \,.
\ee
The latter operator is a linear endomorphism of the $2^{16}$-dimensional quantum spinor space.
We note in passing that the validity of the assumptions made here (and the validity of the
final result Eq. \eqref{Ralphaq}) has been explicitly checked in Ref. \cite{Damour:2017cpi} in
the case of the (Bianchi IX) $D=4$, $N=1$, supergravity model.

Using again the simple functorial nature of Fock quantization, we can finally write down some
of the relations satisfied both by the classical, and the quantum, reflection operators
$\mathcal{ R}_{\alpha_I}$. Let us recall that, motivated by the structure of the fermionic billiards
arising in the near-singularity behavior of supergravity, Ref. \cite{Damour:2009zc} introduced,
when working within specific finite-dimensional representations of the maximally compact subalgebras of
physically relevant hyperbolic Kac-Moody algebras (namely $K[E_{10}] \subset E_{10}$,
and  $K[AE_3] \subset AE_3$) the notion of  spin-extended Weyl groups, generated
by fermion reflection operators associated with the simple roots $\alpha_i$ of the
considered Kac-Moody algebra, say $G$.  See  Ref. \cite{Koehl} for 
a mathematical definition of spin-extended Weyl groups (for general simply-laced Kac-Moody algebras)
as a part of the definition of spin-covers of  maximal compact Kac-Moody subgroups. 

As here we are in a setting where we constructed finite-dimensional representations (for a non-simply-laced case)
of $K(G_2^{++})$, we can define spin-extensions of the Weyl group of $G_2^{++}$ as the group
of linear operators generated by (to be explicit)
\be\label{Ralphasimple}
\mathcal{ R}_{\alpha_i} = e^{ i  \frac{\pi}{2}   J_{\alpha_i}} \,,
\ee
where $i$ labels the simple roots (in our case $i=(1), (12), (23), (34)$), and where the linear
operator $ J_{\alpha_i}$ is taken in one of the finite-dimensional representations defined above.
Specifically, we can take $ J_{\alpha_i}$ in the 16-dimensional vector-spinor representation
Eq. \eqref{defJ} (corresponding to the classical reflection operators \eqref{RalphaG}), or
in the $2^{16}$-dimensional quantum vector-spinor representation defined in Eq.  \eqref{defJq}.

The last point we wish to make here is that, in both these representations, the four reflection
operators, $r_i = \mathcal{ R}_{\alpha_i}$, listed in Eq. \eqref{Ralphasimple}, associated with the four simple roots of $G_2^{++}$,
satisfy the  following generalized Coxeter relations 
 \be \label{r8}
 r_i^8=1;
 \ee 
 together with the ``braid relations" (see Refs.\cite{KacPeterson85,Koehl})
 \be\label{braid}
 r_i r_j r_i \cdots = r_j r_i r_j \cdots \, {\rm with} \, m_{ij} \, {\rm factors \, on \, each\, side} \, .
 \ee
 Here, $i$, and $j$, with $i\neq j$, are labels for the nodes of the  Dynkin diagram of the considered Kac-Moody group.
 The positive integers $m_{ij}$ entering the braid relation \eqref{braid} are defined from the corresponding values of the nondiagonal elements of the Cartan matrix $a_{ij}$ (which are  negative integers, while $a_{ii}=2$).
Namely (see \cite{KacPeterson85})
\be
m_{ij} = \left\{ 2,3,4,6,0 \right\} \, \, {\rm if} \, \, a_{ij}a_{ji}=  \left\{ 0,1,2,3, \geq 4 \right\} \, \,( {\rm respectively }) \, .
\ee
Note that in our case the values $i=(1), j=(12)$ have $a_{ij}a_{ji}=3$, corresponding to $m_{ij}=6$. In that case
the braid relation, Eq. \eqref{braid}, explicitly reads
\be
 r_{(1)} r_{(12)} r_{(1)}  r_{(12)} r_{(1)} r_{(12)} =r_{(12)} r_{(1)} r_{(12)} r_{(1)} r_{(12)}  r_{(1)}
\ee
The validity of  Eq. \eqref{r8} for the $16$-dimensional vector-spinor
classical representation is easily checked to follow from the half-integral nature of the eigenvalues of the 
basic gamma matrices $\frac{i \gamma^{ a b}}{2}$ and $\frac{ \gamma^{ a}}{2}$
entering their definitions. Indeed, let us look again at the definition of the
classical action of $J_{\alpha}$ in the $16$-dimensional vector-spinor representation
\be \label{Jalphacl}
\left( J_{\alpha} . \Phi \right)^{aA} \equiv (\delta^a_b - 2 \alpha^a \alpha_b) \left(J^s_{\alpha}\right)^A_{\ B} \Phi^{bB} \,, \
\ee
where $J_{ab}^s= i\frac{ \gamma^{ab}}{2}$ while $J_{a}^s= \frac{ 3 \gamma^{a}}{2}$.

The eigenvectors $v^{aA}$ of $J_{\alpha}$ can be looked for in factorized form, namely
$v^{aA}= v^a \xi^A$ where $v^a$ is an eigenvector of the matrix $\delta^a_b - 2 \alpha^a \alpha_b$
(say $(\delta^a_b - 2 \alpha^a \alpha_b) v^b= \lambda_v v^a$) while $\xi^A$ is an eigenvector
of the spin part $J^s_{\alpha}$ (say $\left(J^s_{\alpha}\right)^A_{\ B} \xi^B= \lambda_s \xi^B$).
The eigenvalue of $ J_{\alpha}$ corresponding to $v^{aA}= v^a \xi^A$ is equal to the product
$  \lambda_{vs} =  \lambda_v  \lambda_s$. 
The four eigenvectors of  $\delta^a_b - 2 \alpha^a \alpha_b$ are: (i) any vector parallel to $\alpha^a$,
with eigenvalue $1- 2 \langle \alpha \alpha \rangle$; and (ii) three vectors orthogonal to $\alpha^a$,
with eigenvalue 1. Using the fact that the squares of the matrices $ i \gamma^{ab}$ and $\gamma^a$ are
equal to the unit matrix, one finds that the four eigenvalues of $J_{ab}^s= i\frac{ \gamma^{ab}}{2}$ are 
$\{+\frac12, +\frac12, -\frac12,-\frac12 \}$, while the four eigenvalues of  $J_{a}^s= \frac{ 3 \gamma^{a}}{2}$ are
$\{+\frac32, +\frac32, -\frac32,-\frac32 \}$. 
 Using the fact that the squared roots $\langle \alpha \alpha \rangle= G^{ab}  \alpha_a \alpha_b$ are
equal to $2$ for the long symmetry roots $\alpha^{(ab)}$, but equal to $\frac23$ for the
short electric roots, one finds that the corresponding vector eigenvalues $\lambda^{(ab)}_v$'s  are $\{-3,1,1,1\}$, 
while the $\lambda^{(a)}_v$'s are  $\{- \frac13,1,1,1\}$. As a consequence the sixteen product eigenvalues
 $  \lambda_{vs} =  \lambda_v  \lambda_s$ have the values 
 $\{\pm \frac32,\pm \frac12,\pm \frac12,\pm \frac12,\}$ for the symmetry walls, and the values 
  $\{\pm \frac12,\pm \frac32,\pm \frac32,\pm \frac32,\}$ for the electric walls. [Note the cancellation of the
  $\frac13$ coming from the anomalous  $\langle \alpha \alpha \rangle= \frac23$ by the extra factor $C^s=3$
  in the definition of  $J_{a}^s$.]
  
  When passing from the $16$-dimensional classical-gravitino representation to the  $2^{16}$-dimensional
  quantum-gravitino representation, the explicit forms of the action of ${\widehat J}_{\alpha}$ at various
  fermion levels (see Eqs. \eqref{Jlevel2}, \eqref{Jlevel3}) show that the eigenvalues at level $N_F=n$ are given by sums 
  $\lambda^{N_F=n}= \lambda_1 + \lambda_2 + \cdots + \lambda_n$, corresponding to a factorized
  eigenvector $v_1\wedge v_2 \wedge \cdots \wedge v_n$, where each $v_p$ is itself of the
  factorized form $v_p^{aA}= v_p^a \xi_p^A$ (under the condition that these wedge products
  do not vanish). [The full spectrum of the ${\widehat J}_{\alpha}$'s, with their multiplicities, will be found in Appendix \ref{Jspectrum}.]
  This result immediately shows that all the eigenvalues of ${\widehat J}_{\alpha}$
  will be half-integral (or integral). This guarantees that the 8th power of
  ${\widehat {\mathcal R}}_{\alpha} = e^{ i  \frac{\pi}{2}  \widehat J_{\alpha}}$
  is equal to 1.

We have verified the validity of the braid relations \eqref{braid}
for the classical, $16$-dimensional vector-spinor representation of the $J_{\alpha}$'s
 by a direct computation. For instance,
 \be \label{braidc}
 {{\mathcal R}}_1 { {\mathcal R}}_{12}
 { {\mathcal R}}_1 {{\mathcal R}}_{12} { {\mathcal R}}_1
 { {\mathcal R}}_{12}=  {{\mathcal R}}_{12} { {\mathcal R}}_1
 { {\mathcal R}}_{12} {{\mathcal R}}_1 { {\mathcal R}}_{12}
 { {\mathcal R}}_1\,,
 \ee
 while
 \be
  { {\mathcal R}}_{12}  { {\mathcal R}}_{23}  { {\mathcal R}}_{12}= { {\mathcal R}}_{23} { {\mathcal R}}_{12} { {\mathcal R}}_{23}\,.
 \ee
 These results can then be lifted to the full
 $2^{16}$-dimensional quantum-gravitino representation by using the functorial nature of
 the Fock-representation expressions  Eqs. \eqref{Jlevel2}, \eqref{Jlevel3}. Indeed, they imply that
 corresponding exponentiated operators, 
  ${\widehat X}_{\alpha} = e^{ x  \widehat J_{\alpha}}$, act, when considered
  at any given level\footnote{$\widehat J_{\alpha}$ commutes with $\widehat N_F$ and 
  therefore any function of  $\widehat J_{\alpha}$ acts within any fixed-$N_F$ space.},
   $N_F$, as a product of corresponding classical exponentiated factors. E.g., at level 2,
   we have
   \be
   e^{ x  \widehat J_{\alpha}}( u\wedge v )=  ( e^{ x   J_{\alpha}} u) \wedge (   e^{ x   J_{\alpha}}  v).
   \ee
Such a general product action applies in particular to the reflection operators  ${\widehat {\mathcal R}}_{\alpha} = e^{ i  \frac{\pi}{2}  \widehat J_{\alpha}}$, and thereby also to
the relevant braid operators which are made of products of  ${\widehat {\mathcal R}}_{\alpha}$'s.
 As a consequence, the equality of two braid classical combinations, e.g. Eq. \eqref{braidc}, entails
 the equality of the corresponding quantum combination at all levels, so that, e.g.,
 \be \label{braidq}
 {\widehat {\mathcal R}}_1 {\widehat {\mathcal R}}_{12}
 {\widehat {\mathcal R}}_1 {\widehat {\mathcal R}}_{12} {\widehat {\mathcal R}}_1
 {\widehat {\mathcal R}}_{12} = {\widehat {\mathcal R}}_{12} {\widehat {\mathcal R}}_1
 {\widehat {\mathcal R}}_{12}  {\widehat {\mathcal R}}_1 {\widehat {\mathcal R}}_{12}
 {\widehat {\mathcal R}}_1 \,,
 \ee
holds in the $2^{16}$-dimensional quantum-gravitino representation.

%%%%%%%
\section{Conclusions}

Let us summarize our main results on the supersymmetric quantum dynamics of the cosmological models obtained 
by reducing $D=5$ supergravity to one timelike dimension, i.e. by considering the consistent truncation where
the spatial metric, $h_{ij}$, the vector potential, $A_i$, and the spatial components of the gravitino, $\psi^{i A}$,
depend only on time. 

(1) We constructed a consistent quantization of this model, with the fourteen bosonic coordinates
quantized \`a la Schr\"odinger ($p = \frac{\hbar}{i} \frac{\partial}{\partial q}$), while the suitably
redefined spatial gravitino field $\Phi^{aA}= (\det h)^{\frac14}\gamma^a \theta^a_i \psi^{iA}$
satisfies simple anticommutation relations 
$\{ \widehat \Phi^{aA} , {\widehat \Phi}^{\dagger bB} \} = \hbar G^{ab} \, \delta^{AB}$.
%(which characterize a Spin(16,16) Clifford algebra).
Here, $G^{ab}$ is the inverse of the metric $G_{ab}$ in the Cartan space of $G_2^{++}$:
$G_{ab} \dot\beta^{ a} \dot\beta^{ b} \equiv \sum_a(\dot\beta^{ a})^2-(\sum_a\dot\beta^{ a})^2$,
where the $\beta^a$'s are the logarithmic scale factors of the spatial metric $h_{ij}$, see Eq. \eqref{iwasawa}.
In other words, the wave function of the Universe  is a $2^{16}$--component spinor of  Spin(24,8)
which depends on the fourteen bosonic configuration variables $h_{ij}$, $A_i$ (with $i=1,2,3,4$).
The latter variables are usefully replaced by the four logarithmic scale factors, $\beta^a$, the six 
off-diagonal Iwasawa variables  $\N^a_{\ \  i}$ (with $a <i$), and the four electric variables 
$ B_a \equiv A_i({\cal N}^{-1})^i_{\ a}$.

(2) Quantum states  $\vert{\bm\Psi }\rangle$ are described  by wavefunctions 
${\Psi}_{\sigma} (\beta^a , \N^a_{\ \  i}, B_a)$ (where the spin index $\sigma$ takes $2^{16}$ values)
that must satisfy the eight (Dirac-like) supersymmetry constraints
$\widehat {\cal S}_A \, \vert{\bm\Psi}\rangle = 0 , \widehat { {\cal S}}^{\dagger}_A  \, \vert{\bm\Psi}\rangle = 0$,
as well as the Hamiltonian constraint $\widehat H \, \vert{\bm\Psi}\rangle = 0$. We have checked  the consistency
of the algebra of constraints (see Eqs. \eqref{qidentities1}, \eqref{qidentity2}, \eqref{qidentity3}) 
when using an ordering ensuring an up-down symmetry in fermion space (i.e. symmetry under swapping
$ \Phi \leftrightarrow \Phi^{\dagger}$).

(3)  The hyperbolic Kac-Moody algebra $G_2^{++}$ shows up in the bosonic sector in the fact that
the bosonic part of the Hamiltonian describes a null geodesic over the symmetric space $G_2^{++}/K(G_2^{++})$
when setting to zero some higher-level Kac-Moody terms formally corresponding to some spatial gradient
terms on the supergravity side \cite{Mizoguchi:2005zf}. The root structure of  $G_2^{++}$ is reflected
in the bosonic Hamiltonian through the presence of a Toda-like structure:
\be \label{H0bis}
H^{(0)}= G^{ab} \pi_a \pi_b + 2 \sum_{a<b} e^{-2(\beta^{ b}-\beta^{ a})} ({P}_{ab})^2+ \frac12 \sum_a e^{-2\,\beta^a}(P^a)^2\,,
\ee
where $\pi_a$ is the conjugate momentum to $\beta^a$,
$P^a$ is the momentum conjugate to $B_a$, and where ${P}_{ab}$ is a momentumlike variable associated with
$\N^a_{\ \  i}$. Here $\alpha^{(ab)}(\beta)= \beta^b-\beta^a$, and $\alpha^{(a)}(\beta)$ are
linear forms in the $\beta$'s which correspond to (real) roots of $G_2^{++}$. In particular,
they feature the four simple roots $\alpha_1=\alpha^{(1)}$,  $\alpha_2=\alpha^{(12)}$, 
 $\alpha_3=\alpha^{(23)}$,  $\alpha_4=\alpha^{(34)}$ defining the Cartan matrix, Eq.~\eqref{Cartan},
 of $G_2^{++}$.

(4) The $K(G_2^{++})$ structure associated with the fermions shows up in the fermionic sector in several
 ways. The part $ \widehat H^{(2)}$ of the quantum Hamiltonian that is bilinear in the fermions reads
 \be \label{qH2ter}
\widehat H^{(2)}= + 2 \sum_{a<b} e^{-(\beta^{ b}-\beta^{ a})}\widehat {P}_{ab} \widehat J_{ab}(\Psi) - \frac{1}{\sqrt{3}}\sum_a e^{-\beta^a}\widehat P^a \widehat J_a(\Psi)\,.
\ee
This fermion-quadratic contribution associates to each one
of the wall roots, $\alpha_I(\beta)=(\alpha^{(ab)}(\beta), \alpha^{(a)}(\beta))$, entering the
bosonic Hamiltonian, a corresponding fermion bilinear $\widehat J_{\alpha_I}$. 
The latter quantum fermion bilinears generate a $2^{16}$-dimensional representation of $K(G_2^{++})$.
Indeed, the four operators $\widehat J_{\alpha_i}$, $i=1,2,3,4$, corresponding to the four
simple roots of  $G_2^{++}$, satisfy the Serre-Berman relations, Eqs. \eqref{berman}.

(5) In the short-wavelength limit, the propagating-wave solutions of the constraints that
exist in the middle of the fermionic Fock space ($4 \leq N_F \leq 12$) transform, upon reflection 
on each of the (symmetry or electric) potential wall delimiting the boundary of the billiard
chamber (identified with the Weyl chamber of $G_2^{++}$), under the corresponding four
quantum reflection operators
\be\label{Ralphaqbis}
{\widehat {\mathcal  R } }_{\alpha_i} = e^{  i  \frac{\pi}{2}  \widehat J_{\alpha_i}} \,.
\ee
These quantum reflection operators satisfy the generalized Coxeter relations given in Eqs. 
\eqref{r8}, \eqref{braid}. These relations define a spinorial extension of the Weyl group of $G_2^{++}$.

(6) The quartic-in-fermion contribution to the quantum Hamiltonian $ \widehat \mu^2 \equiv\widehat H_1^{\prime (4)}$
(as defined in Section \ref{qconstraints}) satisfy two remarkable $K(G_2^{++})$-related properties:
First, it is invariant under the generators $\widehat J_{\alpha_i}$, $i=1,2,3,4$ of $K(G_2^{++})$.
Second, it happens to be expressible (see Eq. \eqref{mu2}) in terms of two $K(G_2^{++})$-invariant {\it fermion bilinears}, the
(sesquilinear) fermion number, $\widehat N_F \equiv G_{ab} \widehat \Phi^{a  \dagger}  \widehat \Phi^{b }$,
and the bilinear $\widehat C_F \equiv G_{ab} \widehat \Phi^{a A} C_{AB} \widehat \Phi^{b B}$,
which involves the ``charge conjugation" matrix $ C_{AB}$ of the (spatial) 
$\gamma_i$ matrices ($C \gamma_i C^{-1}= - \gamma_i^T$).

The invariance of $\widehat N_F $ and $\widehat C_F$ under the 
 $\widehat J_{\alpha_i}$'s stems from the invariance of the two  corresponding  bilinear forms
$H(\Phi_1, \Phi_2) =G_{ab}  \Phi_1^{a A  \dagger}  \delta_{AB} \Phi_2^{b  B}$ 
and $ J(\Phi_1, \Phi_2) =G_{ab}  \Phi_1^{a A  }  C_{AB} \Phi_2^{b  B}$ under the action
of the generators of $K(G_2^{++})$ in the  16-dimensional space
defined by the  (classical) vector-spinor representation. [Here, we consider this representation
from a mathematical point of view, i.e. within the vector space of complex-valued vector-spinors
$ \Phi^{a A  }$.  The quantum representation being correspondingly built by Fock quantization,
as discussed in Section \ref{KM}.]  The sesquilinear form $H$ is {\it hermitian}, with signature $(12^+,4^-)$,
while the bilinear form $J$ is {\it symplectic}. The fact that the representatives of the generators of the
(infinite-dimensional) algebra $K(G_2^{++})$ within this 16-dimensional representation leave invariant these two forms indicate that the image of $K(G_2^{++})$ within this representation is the intersection of the
Lie algebra of the pseudo-unitary group $U(12,4)$ and of the
symplectic group $Sp(16, {\mathbb C})$. By definition, this intersection is the Lie algebra
$sp(6,2)$ of the quaternionic pseudo-unitary group $Sp(6,2)$. The image Lie algebra $sp(6,2)$ 
of Lie$[K(G_2^{++})]$ has
real dimension 136 (as can be directly checked by looking at the general solution of the invariance
conditions, $u^{\dagger} H + H u=0$, $u^T J + J u=0$, of $H$ and $J$ under an
infinitesimal $GL(16, {\mathbb C})$ transformation $\delta \Phi = u \cdot \Phi$).

Our results  open new perspectives that we hope to explore in future work.  
The most promising one is that
 our finding that the quartic-in-fermions term commutes with the generators
of the involutory subalgebra  $K(G_2^{++})$ (which generalizes the similar property found for
 $\N=1$, $D=4$ supergravity \cite{Damour:2014cba}), raises the hope that such a property will also
 hold for the maximal supergravity $\N=8$ in $D=4$, or $\N=1$ in $D=11$, i.e. that the quartic-in-fermions term
 in the Hamiltonian is invariant under $K(E_{10})$. Let us note in this respect that, as shown by Eq. \eqref{H4},
 the quartic-in-fermions term in the Hamiltonian is the sum of the original fermion-quartic part of the Lagrangian
 and of a sum of the squares of the $J_{\alpha_I}$ bilinears corresponding to the roots explicitly appearing
 in the Lagrangian (the latter terms being generated by the Legendre transform associated to the 
 velocity dependence of the couplings of $\N^a_{\ \ i}$ and $A_i$ to fermions). Though our analysis has
 truncated away the couplings to spatial derivatives, it has retained all the velocity-dependent couplings.
 [And, a similar analysis can be implemented for $D=11$ supergravity.]
 This suggests that the value of $\mu^2$ obtained in such one-time-dimension reductions is relevant to the exact
 supergravity dynamics. Therefore, finding a value of  $\mu^2$ that is invariant under the relevant
 involutory algebra is a strong signal of a hidden Kac-Moody-related symmetry. [In previous
 Kac-Moody-coset analyses, it was argued that $\mu^2$ is naturally given by a quadratic Casimir, 
 $\frac12 \sum_{\alpha} J_{\alpha}^2$, involving a formal sum over the infinite number of positive 
 roots \cite{deBuyl:2005zy,Damour:2014cba}.]
  It will be therefore important to see whether an extension of our analysis to the $D=11$-supergravity 
  fermion sector  leads to a $\mu^2$ that is invariant under $K(E_{10})$.
  
  If this  is the case, besides being a clear confirmation of a hidden $K(E_{10})$ symmetry,
   it will also probably imply that 
 $ \widehat \mu^2 \equiv\widehat H_1^{\prime (4)}$ is a c-number, rather than a fermionic operator,
 because we have checked that there are no non-trivial $K(E_{10})$-invariant
 (symplectic) bilinears of the type, $C_F=G_{ab} \Phi^{a}C \Phi^{b} =G_{ab} \Phi^{aA}C_{AB} \Phi^{bB}$,
 that allowed expressions of the type Eq. \eqref{mu2} to exist. [In $D=4$, $\widehat \mu^2$ was
 quadratic in $C_F$ with $C= \gamma_5=\gamma^{ 0} \, \gamma^{ 1} \, \gamma^{ 2} \, \gamma^{ 3}$, while
 in $D=5$, $C_F$ involved the  spatial charge conjugation matrix. We recall that, like in $D=4$,  the gravitino
 is a Majorana spinor in $D=11$.]

%%%%%%%%%
\begin{acknowledgments}
We thank Axel Kleinschmidt and Antoine van Proeyen for informative exchanges.
%Ph. S. thanks Fran{\c c}ois Englert for enlightening discussions about Clifford algebra structures.
Ph. S. acknowledges the hospitality and the stimulating environment of the
Institut des Hautes Etudes Scientifiques. 
%his work has been partially supported by the PDR ``Gravity and extensions'' from the F.R.S.-FNRS (Belgium) (convention T.1025.14).
\end{acknowledgments}

\appendix

\section{Spectrum of the quantized $J$ operators} \label{Jspectrum}

\begin{widetext}

\subsection*{ $\widehat J_{[ab]}$ spectrum}
\begin{equation}
\begin{aligned} N_F=0,\, 16\qquad&\lambda = \quad 0\big\vert_{1} \ &, &\ &\ &\ &\ &\ \\
N_F=1,\, 15:\qquad&\lambda = \pm \frac 32\big\vert_{2} \ &,&\  \pm \frac 12\big\vert_{6}&\ &\ &\ &\ \\
 N_F=2,\, 14:\qquad&\lambda = \pm 3\big\vert_{1} \ &,&\   \pm 2 \big\vert_{12} \ &,&\  \pm 1 \big\vert_{27} \ &,& \quad  0 \big\vert_{40}&&\\
N_F=3,\, 13:\qquad&\lambda = \pm \frac72\big\vert_{6} \ & ,&  \pm \frac{5}2 \big\vert_{36} \ & ,&\ \pm \frac32\big\vert_{94} \ &,& \pm \frac12\big\vert_{144} &\ \\
N_F=4,\, 12:\qquad&\lambda =\pm 4\big\vert_{15}  &,&\  \pm 3\big\vert_{76}\ &, & \  \pm 2\big\vert_{222}\ &,&\  \pm 1\big\vert_{372}\ &,&\quad  0\big\vert_{450}&\ \\
N_F=5,\, 11:\qquad&\lambda =\pm\frac 92\big\vert_{20}\quad &,&\  \pm\frac72\big\vert_{120}\quad&,&\quad\pm \frac52\big\vert_{366}\ &,&\ \pm \frac32\big\vert_{712}\ &,&\ \pm \frac 12\big\vert_{966}\\
N_F=6,\, 10:\qquad&\lambda =\pm 5\big\vert_{15}\  &,&\  \pm 4\big\vert_{132}\ &,&\ \pm 3\big\vert_{466}\ &,&\ \pm 2\big\vert_{966}\ &,&\ \pm 1\big\vert_{1527}\ &,&\quad 0\big\vert_{1776}\\
N_F=7,\, 9 : \qquad&\lambda =\pm\frac {11}2\big\vert_{6}\  &,&\  \pm\frac92\big\vert_{92}\ &,&\  \pm\frac72\big\vert_{402}\ &,&\ \pm \frac 52\big\vert_{1020}\,&,&\ \pm \frac 32 \big\vert_{1812}\ &,&\ \pm \frac 12\big\vert_{2388}\\
N_F=8:\qquad&\lambda = \pm 6\big\vert_{1}\ & ,&\pm5\big\vert_{36}\ & ,&\quad \pm 4\big\vert_{249}\ &,&\ \pm 3\big\vert_{764} \ &,&\pm 2\big\vert_{1599} \ & ,&\pm 1\big\vert_{2400}\ &,&\quad\ 0\big\vert_{2772}
 \end{aligned}
\end{equation}
\subsection*{ $\widehat J_{a}$ spectrum}
\begin{equation}
\begin{aligned} N_F=0,\, 16\qquad&\lambda = \quad 0\big\vert_{1} \ &, &\ &\ &\ &\ &\ \\
N_F=1,\, 15:\qquad&\lambda = \pm \frac 32\big\vert_{6} \ &,&\  \pm \frac 12\big\vert_{2}&\ &\ &\ &\ \\
 N_F=2,\, 14:\qquad&\lambda = \pm 3\big\vert_{15} \ &,&\   \pm 2 \big\vert_{12} \ &,&\  \pm 1 \big\vert_{13} \ &,& \quad  0 \big\vert_{40}&&\\
 N_F=3,\, 13:\qquad&\lambda = \pm \frac92\big\vert_{20} \ & ,&  \pm \frac{7}2 \big\vert_{30} \ & ,&\ \pm \frac52\big\vert_{36} \ &,&\  \pm \frac 32 \big\vert_{144} \ &,& \pm \frac12\big\vert_{80} &\ \\
N_F=4,\, 12:\qquad&\lambda =\pm 6\big\vert_{15}  &,&\  \pm 5\big\vert_{40}\ &, & \  \pm4\big\vert_{55}\ &,&\  \pm 3\big\vert_{180}\\
&\phantom{\lambda =}\pm  2\big\vert_{207}\ &,&\quad  1\big\vert_{228}\ &,&\ 0\big\vert_{370}\\
N_F=5,\, 11:\qquad&\lambda =\pm\frac {15}2\big\vert_{6}\quad &,&\  \pm\frac{13}2\big\vert_{30}\quad&,&\quad\pm \frac{11}2\big\vert_{50}\ &,&\ \pm \frac92\big\vert_{170}\\&\phantom{\lambda =}\pm \frac 72\big\vert_{290}\ &,&\ \pm \frac 52\big\vert_{360}\ &,&\ \pm \frac 32\big\vert_{666}\ &,&\ \pm \frac 12\big\vert_{612}\\
N_F=6,\, 10:\qquad&\lambda =\pm 9\big\vert_{1}\  &,&\  \pm 8\big\vert_{12}\ &,&\ \pm 7\big\vert_{27}\ &,&\ \pm 6\big\vert_{96}\ &,&\ \pm 5\big\vert_{235}\ &,& \pm 4\big\vert_{340}\\
&\phantom{\lambda =}\pm {3}\big\vert_{720}\  &,&\  \pm 2\big\vert_{900}\quad&,&\quad\pm 1\big\vert_{1005}\ &,&\ 0\big\vert_{1336}\\
N_F=7,\, 9 : \qquad&\lambda =\pm\frac {19}2\big\vert_{2}\  &,&\  \pm\frac {17}2\big\vert_{8}\ &,&\  \pm\frac {15}2\big\vert_{30}\ &,&\ \pm \frac{13}2\big\vert_{108}\,&,&\ \pm \frac {11}2\big\vert_{192}\ &,&\ \pm \frac 92\big\vert_{470}\\
&\phantom{\lambda =}\pm\frac {7}2\big\vert_{780}\  &,&\  \pm\frac {5}2\big\vert_{990}\ &,&\  \pm\frac {3}2\big\vert_{1590}\ &,&\ \pm \frac{1}2\big\vert_{1550}\\
N_F=8:\qquad&\lambda = \pm 10\big\vert_{1}\ & ,&\pm 9\big\vert_{4}\ & ,&\quad \pm 8\big\vert_{25}\ &,&\ \pm 7\big\vert_{60} \ &,&\pm 6\big\vert_{174} \ & ,&\pm 5\big\vert_{396}\\
&\phantom{\lambda = }\pm 4\big\vert_{585}\ & ,&\pm 3\big\vert_{1140}\ & ,&\quad \pm 2\big\vert_{1425}\ &,&\ \pm 1\big\vert_{1600}  \ &,&\quad\ 0\big\vert_{2050}
 \end{aligned}
\end{equation}

\end{widetext}

\section{Explicit form of the supersymmetry constraints} \label{susyex}

In this Appendix we use a slightly different notation from the one used in the text. Vector indices
are denoted $a, b, \cdots$ (as in the text), while spinor indices are denoted $\alpha, \beta, \cdots$. 
The composite indices
combining these two types of indices (denoted ${\cal A} = a A$ in the text) are denoted here as
$A = a \alpha$. When a spinor, or a composite index, belongs to some $\Phi^{\dagger}$ we dot it
to indicate its origin, e.g.  
$ ({\hat \Phi}^{a \alpha})^\dagger={\hat \Phi}^{\dagger a \dot \alpha}=({\hat \Phi}^A)^\dagger={\hat \Phi}^{\dagger\dot A}$. The right-handside $G^{ab} \, \delta^{\alpha \beta}$ (with $\hbar=1$)
of the third (non trivial) anticommutation relations Eq. \eqref{CRPhi} is denoted $\Delta^{A\dot B}$, i.e.
\be
\{{\hat \Phi}^A,{\hat \Phi}^{\dagger\dot B}\}=\Delta^{A\dot B}\,.
\ee
Though, with our normalization $\Delta^{A\dot B}= G^{ab} \, \delta^{\alpha \beta}$ is real
and symmetric, it is useful (for keeping track of hermitian conjugations in contracted indices)
to  denote its complex conjugate as $(\Delta^{A\dot B})^\star=\Delta^{B\dot A}$.
We also denote the (purely imaginary) numerical coefficients  $  c^{\alpha}_{  {\cal A} {\cal B}  {\cal C}}= -  c^{\alpha}_{  {\cal C} {\cal B}  {\cal A}}$ entering
Eq. \eqref{qS3Phi} as $\sigma^{\alpha}_{  {\dot B}  [A  C]}= - \sigma^{\alpha}_{  {\dot B}  [C  A]}$,
so that $ {\cal S}^{\alpha}$ reads
\be
\mathcal S^\alpha=-i\,\partial_{\beta^a}{\hat \Phi}^{a\alpha}+\sigma^\alpha_{\ \dot P[BC]}{\hat \Phi}^B{\hat \Phi}^{\dagger\dot P}{\hat \Phi}^C\,.
\ee
The following contraction of the $\sigma$ coefficients plays a distinguished role:
\be
\sigma^\alpha_{\ \dot P[BC]}\Delta^{B\dot P}=-i\,\nu_c\,\delta^{\alpha}_{\gamma}\,.
\ee
Here $C=c\gamma$ and the four components of the vector $\nu$
are  $\nu_a=-\frac1 4\{19,16,13,10\}$.

With this notation, the explicit form of the supersymmetry constraint ${\cal S}^\alpha \vert X\rangle =0$,
when acting on a plane-wave state of fermion level $k$ written as
\be
\vert X\rangle=e^{i\,\pi_a\beta^a}\,X_{\dot A_1\cdots  \dot A_k} {\hat \Phi}^{\dagger \dot A_1}\cdots  {\hat \Phi}^{ \dagger \dot A_k}\vert 0\rangle_-\,,
\ee
reads
\begin{widetext}
\be
k\,\Big( ({\pi}_a-i\,\nu_a)\Delta^{a\alpha \dot P}X_{\dot P\dot\,\dot A_1\cdots  \dot  A_{k-1}}- (k-1)\,\sigma^\alpha_{\ [\dot A_1\vert [BC]  }\Delta^{C\dot P}\Delta^{B\dot Q}\,X_{\dot P\dot Q\vert \dot A_2\cdots  \dot A_{k-1}]}\Big)=0\,.
\ee
The corresponding explicit form of the constraint  ${\cal S}^{\dagger \dot \alpha} \vert X\rangle =0$ reads
\be
\Big((\pi_{a_1} + i\,\nu_{a_1})\delta^{\dot \alpha}_{\dot \alpha_1}X_{\dot A_2\cdots \, \dot  A_{k+1}}+k\,\sigma^{* \dot\alpha}_{\ \  B \dot A_1\dot A_2 }\Delta^{B\dot P} \,X_{ \dot P  \dot A_3\cdots  \dot A_{k+1}}\Big )_{[\dot A_1\dot A_2 \cdots \dot A_{k+1}]}=0\,,
\ee
where the last subscript indicates antisymmetrization with respect to the composite indices
$\dot A_1\dot A_2 \cdots \dot A_{k+1}$ (with $\dot A_1= a_1 \dot \alpha_1$).

\end{widetext}

\end{document}